\documentclass[final,3p,times]{elsarticle}

\usepackage{graphicx}

\usepackage[separate-uncertainty=true]{siunitx}

\usepackage{amssymb}
\usepackage{bbold}

\usepackage{amsmath}
\usepackage[table, x11names]{xcolor}
\colorlet{lightgrayX50}{lightgray!50}

\usepackage{gensymb}        

\usepackage[para,online]{threeparttable}

\usepackage{makecell}

\usepackage{multirow}
\usepackage{tabto}

\usepackage[draft]{hyperref}
\usepackage[capitalise]{cleveref}

\usepackage{booktabs}

\usepackage[]{algorithm2e}

\usepackage{color,soul}

\usepackage{lineno}

\usepackage{natbib}

\newcommand{\Image}{\mathcal{I}}
\newcommand{\argmax}{\mathop{\mathrm{argmax}}}    
\newcommand{\argmin}{\mathop{\mathrm{argmin}}}    

\biboptions{square,sort&compress}

\journal{Materials Characterization}

\begin{document}
\begin{frontmatter}

\title{Accurate reconstruction of EBSD datasets by a multimodal data approach using an evolutionary algorithm}

\address[label1]{Materials Department, University of California, Santa Barbara, CA 93106, USA}
\address[label2]{Univ. Lille, CNRS, Centrale Lille, Inria, UMR 9189 - CRIStAL, F-59000 Lille, France}

\author[label1]{Marie-Agathe Charpagne\corref{cor1}}
\cortext[cor1]{corresponding author}
\ead{marie.charpagne@engineering.ucsb.edu}
\author[label2]{Florian Strub}
\author[label1]{Tresa M. Pollock}

\begin{abstract}
A new method has been developed for the correction of the distortions and/or enhanced phase differentiation in Electron Backscatter Diffraction (EBSD) data. Using a multi-modal data approach, the method uses segmented images of the phase of interest (laths, precipitates, voids, inclusions) on images gathered by backscattered or secondary electrons of the same area as the EBSD map. The proposed approach then search for the best transformation to correct their relative distortions and recombines the data in a new EBSD file. Speckles of the features of interest are first segmented in both the EBSD and image data modes. The speckle extracted from the EBSD data is then meshed, and the Covariance Matrix Adaptation Evolution Strategy (CMA-ES) is implemented to distort the mesh until the speckles superimpose. The quality of the matching is quantified via a score that is linked to the number of overlapping pixels in the speckles. The locations of the points of the distorted mesh are compared to those of the initial positions to create pairs of matching points that are used to calculate the polynomial function that describes the distortion the best. This function is then applied to un-distort the EBSD data, and the phase information is inferred using the data of the segmented speckle. Fast and versatile, this method does not require any human annotation and can be applied to large datasets and wide areas. Besides, this method requires very few assumptions concerning the shape of the distortion function. It can be used for the single compensation of the distortions or combined with the phase differentiation. The accuracy of this method is of the order of the pixel size. Some application examples in multiphase materials with feature sizes down to 1 $\mu$m are presented, including Ti-6Al-4V Titanium alloy, Rene 65 and additive manufactured Inconel 718 Nickel-base superalloys.

\end{abstract}


\end{frontmatter}

\sisetup{scientific-notation = true}
\section{Introduction}
\label{sec:introduction}

\subsection{Limits of phase differentiation by EBSD}

Electron Backscattered Diffraction (EBSD) is a powerful tool for gathering local crystallographic data in multiphase materials. This technique enables a fast mapping of microstructures with a good angular resolution (typically 0.5$\degree$), based on the collection of diffraction patterns and the analysis of Kikuchi bands on those patterns. Initially implemented in scanning electron microscopes (SEM), it has been adapted to transmission microscopy (TEM) under the name of Transmission Kikuchi Diffraction (TKD) \cite{Suzuki2013,Echlin2015a}. In the past years, tridimensional EBSD (3D EBSD) by serial sectioning has also emerged as a new tool to reconstruct and analyze materials in three dimensions \cite{Rowenhorst2006,Spanos2008,Lin2010,Calcagnotto2010,Holzer2011,Polonsky2018}.
To index the collected patterns, most commercial software use algorithms based on Hough transformations to detect the characteristic bands on the Kikuchi patterns. This process has the advantage of speed but suffers from some limitations. Because of the diffraction patterns they produce, some phases are intrinsically challenging to index with this method. This is the case for the finely distributed $\beta$ phase in $\alpha - \beta$ Titanium alloys, for example. Another well-known limitation of this method happens in the presence of two phases of similar structures. This is the case of $\gamma$ and $\gamma '$ phases in Nickel and Cobalt-base superalloys. The ordered $L1_{2}$ $\gamma '$ phase produces extra very low-contrast bands compared to the Face Centered Cubic (FCC) $\gamma$, that are undetectable by Hough transforms. In the case of phases producing similar Kikuchi patterns but having significantly different chemistries, one can employ multi-modal data acquisition. Nowell and Wright \cite{Nowell2004} have proposed the simultaneous collection of X‐ray energy dispersive spectroscopy (XEDS) data, along with EBSD data. This method has been shown to enable phase differentiation in complex alloys, and reveal unique crystallographic relationships between phases \cite{West2009,Child2012,Charpagne2016}. However, XEDS-EBSD requires a significant difference in the chemistry of the phases of interest and a minimum number of counts on the XEDS spectra, which can lower the speed of data collection. The acceleration voltage also has to be high enough to enable the detection of the characteristic X-rays emission of the elements of interest. Increasing the acceleration voltage would also result in more counts on the EDS spectra and increase the indexing speed, but is detrimental to the spatial resolution. Indeed, the XEDS interaction volume is about one micrometer at 20 kV in Nickel alloys and increases greatly with the acceleration voltage. \newline
Using the different chemical composition of the phases, other methods involving multi-modal data combination have been implemented. Phases of different composition exhibit different intensities on back-scattered electrons (BSE) images, due to their different Z number. Payton and Nolze \citep{Payton2013} have proposed a method that involves the acquisition of BSE images on a tilted sample, using specific BSE detectors located above the phosphorous screen of the EBSD detector. Those authors have shown that this method enables a more accurate phase differentiation than XEDS-EBSD, since the contrast on the back-scattered images is very sensitive to the variations in chemical composition. They also demonstrate the capability of that method for segmenting small features precisely, with typical dimensions of the order of a few micrometers. The limit of this technique is that not all the EBSD detectors are equipped with such built-in diode sensors. The collection of back-scattered data is usually made by relatively large retractable BSE detectors that have several quadrants, at a 0$\degree$ tilt angle. On the other hand, EBSD data is collected at a 70$\degree$ tilt angle. This, among other reasons, leads to complex distortions in the EBSD versus the BSE data, which makes their precise recombination very challenging.

\subsection{Distortions in EBSD data}
EBSD data is known to suffer from distortions that arise from many instrument and detector artifacts \cite{Nolze2007}. The following types of distortions can be distinguished:
\begin{itemize}
\item First order spatial distortions: those are induced when tilting the sample at 70$\degree$. An area that would be a square at 0$\degree$ tilt turns into a trapezoid if the rotation axis of the stage is not perfectly parallel to the surface of the sample. The borders of the trapeze are also misaligned with those of the non-tilted area. Those distortions can be modeled by a first order affine transformation, involving a translation and a rotation.
\item Second order spatial distortions: the electron beam is deflected by a set of lenses that induce a barrel distortion on the final image. This distortion is mostly visible at low magnification. It can be modeled by a second order polynomial function.
\item Space-time distortions: when a sample is being exposed for some time to the electron beam, charging effects occur. The accumulation of charges on the surface of the sample leads to the deflection of the beam during scanning. This effect occurs in any scanning process. It is usually the most pronounced in the beginning of the scan and becomes more stable as the scan progresses. A good example is shown in \cite{Zhang2014}. The amount of drift is related to the conductivity of the sample itself, the conductivity of the mounting system that is being used, the voltage and aperture, as well as the cleanliness of the scanned surface. The amount of drift is also more important at high than low magnifications \cite{Kammers2013}. There is no simple physical  model to describe the pattern of the drift distortions.
\end{itemize}
Given the complexity of the net resulting distortion, it cannot be easily calculated.


\section{Method}
\label{sec:method}

In this context, we propose a method that enables the compensation of the distortions in EBSD data and adds phase differentiation if wanted, by accurately matching it with BSE images collected at a 0$\degree$ tilt angle, at individual pixel precision and over broad areas. To do so, the Covariance Matrix Adaptation Evolution Strategy (CMA-ES) is used to calculate the relative distortion function between the EBSD data and the BSE image.

\subsection{Working with speckles}
With regard to the recombination of multi-modal data, one first intuitive approach is to use sets of matching points over the area of interest in complementary images. Some methods aiming at correcting the distortions in EBSD maps have been implemented using this principle \cite{Zhang2014}. They require the manual selection of a set of reference and corresponding points in the BSE image and the EBSD data. The higher the number of pairs of points, the greater the precision. However, this step can be time-consuming and does or made difficult by the lack of contrasted features in both EBSD and BSE data. Furthermore, it does not scale up to large or tridimensional datasets, as it may require human annotations for every slice in the dataset. In most 3D EBSD datasets, the process of data collection consists in successive movements of the stage to positions that are practical for machining, imaging, and collection of diffraction patterns. The placement of the stage from a slice to another is never exactly the same \cite{Mingard2014}. This modifies the set of reference points on every slice.\newline
Here, we propose to use speckles of similar features without any human annotations, allowing the recombination of multi-modal data to scale up to large EBSD datasets. To do so, a high number of automatically generated reference points are combined by using microstructural features as speckles, such as pores, a second-phase, or precipitates.
Since the goodness of the superimposition of the speckles can be quantified, it can be used as a parameter to maximize, enabling an automated matching process. Therefore, quantifying the goodness of the superposition of the speckles is critical.

\subsubsection*{Defining a similarity measure between the speckles}
\label{par:measure}

In order to superpose the EBSD and BSE speckles, a quantitative measurement of their similarity is needed. The Dice similarity metric \cite{Dice1945} (also known as F1-score)  describes the relative overlap between the segmented images. More precisely, this score measures the overlap between segmented pixels. Formally, given the binary EBSD speckle $\Image^{EBSD} \in \{0,1\}^{I \times J}$ and the binary BSE image $\Image^{BSE} \in \{0,1\}^{I
\times J}$ where $1$ encodes the segmented object and $0$ the background image, and ${I \times J}$ their domain of definition, the Dice similarity for is computed as follows: 
\begin{align}
	\text{Similarity}(\Image^{EBSD}, \Image^{BSE}) = 2\frac{|\Image_{seg}^{EBSD} \cap \Image_{seg}^{BSE}|}{|\Image_{seg}^{EBSD}| + |\Image_{seg}^{BSE}|}
    \label{eq:similarity}
\end{align}
where $|.|$ encodes the size of the underlying set and $\Image_{seg}^{.}$ only corresponds to the segmented pixels in the image. Thus, $|\Image^{EBSD}_{seg} \cap \Image^{BSE}_{seg}|$ corresponds to the number of matching pixels and $|\Image^{EBSD}_{seg}|$ and $|\Image^{BSE}_{seg}|$ respectively correspond to the total number of pixels in the EBSD and BSE images. This similarity measure counts the number of overlapping segmented pixels normalized by the sum of segmented pixels. It ranges from 0 (no overlapping segmented pixels) to 1 (perfect matching).  The value of the ${Similarity function}(\Image^{EBSD}, \Image^{BSE})$ is referred to as the 'score' in the following. As the Dice similarity is both intuitive, robust and easy to compute, it has been extensively used in several image segmentation applications such as medicine~\cite{taha2015metrics}.

\subsection{Correction work-flow}
\begin{figure}[h]
\centering
\includegraphics[width=0.5\textwidth]{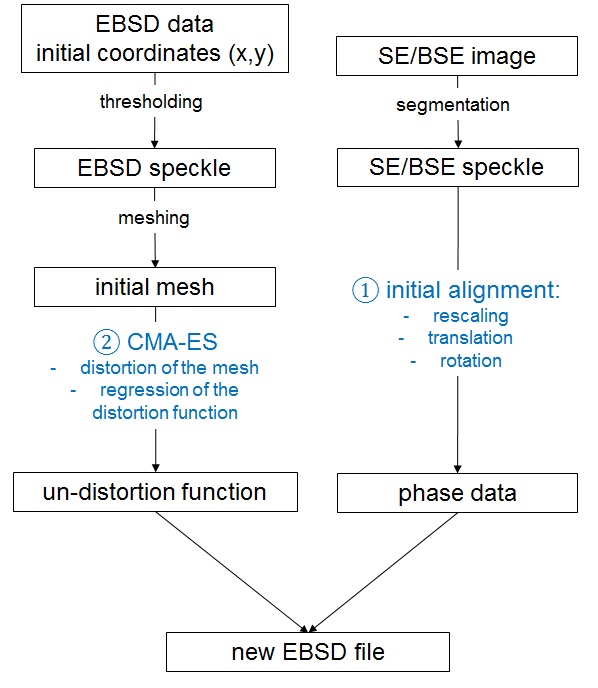}
\caption{Schematic representation of the reconstruction algorithm.}
\label{fig:algo_global}
\end{figure}

The global process of the reconstruction is shown in figure \ref{fig:algo_global}. Starting from the initial EBSD data and the BSE image, two speckles are generated. In the first step (initial alignment), the BSE speckle is rescaled so its pixel size corresponds to that of the EBSD data. The rescaling is done using a nearest neighbor interpolation. The BSE speckle image is also rotated and translated so it is pre-aligned to the EBSD data. To do so, a grid-search over a set of affine transformations is performed, to distort the re-sized speckle over three parameters: two translations along the x and y directions, $t_{x}, t_{y}$, and a rotation of angle $\alpha$. The user is free to choose the widths of the translation and the rotation angle, as well as their increment size. The minimum increment for the translations is one pixel. The step size for the angle can be refined as much as wanted. All the possible translations and rotations within the provided ranges are tested. The transformation leading to the best score is saved and applied to the BSE speckle. The latter is then used in the second step, as the reference speckle for the compensation of the distortions.
In the second step (CMA-ES on fig. \ref{fig:algo_global}), the EBSD speckle is meshed on a regular grid, and the grid is distorted using the CMA-ES optimizer, in an iterative process. The initial and new locations of the points of the grid are used as pairs of control points, to estimate the distortion function. At each iteration, the score produced by the distorted EBSD speckle is calculated. The distortion function corresponding to the distorted speckle that exhibits the best score is used to determine the new location of the points in the new EBSD file and the segmented BSE image is used to fill in the phase data (if needed). The parametrization and calculation of the distortion function are explained in the following.

\subsection{The distortion function}
\label{subsec:distor}

Formally, the goal is to find the distortion function $f$ that maximizes the similarity between the EBSD and the BSE speckles. This leads to the optimization procedure described in eq. \ref{eq:optimisation}:
\begin{align}
	f^* &= \argmax_f \Big( \text{Similarity}(\Image^{EBSD}, \Image^{BSE}) \Big) \\   
	&= \argmax_f  \Big( 2\frac{|f(\Image^{EBSD})_{seg} \cap \Image_{seg}^{BSE}|}{|f(\Image^{EBSD})_{seg}| + |\Image_{seg}^{BSE}| } \Big)
    \label{eq:optimisation}
\end{align}

As described in section~\ref{sec:introduction}, many overlapping distortion phenomena occur while collecting the EBSD data, leading to a non-linear distortion. In the following, the nature of the distortion function is assumed to be polynomial. It is a very weak assumption as polynomial functions can approximate a large range of complex functions. Besides, the proposed method can be easily extended to other function approximators such as Gaussian processes, trees etc. Formally, a polynomial distortion of degree $P$ is defined as follows:

\begin{align}
 f(x,y) =
  \begin{cases}
	x' = \sum_{n=0}^{P} \sum_{k}^{n} c^x_{k,n-k}x^{k}y^{n-k}   \\
	y' = \sum_{n=0}^{P} \sum_{k=0}^{n} c^y_{k,n-k}x^{k}y^{n-k}
\end{cases}
  \label{eq:polynomial_fit}
\end{align}

For instance, given a polynomial fit of degree 2, the coordinates $(x,y)$ of a pixel are transformed into $(x', y')$:

\begin{align}
 f(x,y) =
  \begin{cases}
x' = c^x_{0,0} + c^x_{1,0}x + c^x_{0,1}y + c^x_{1,1}xy + c^x_{2,0}x^2 + c^x_{0,2}y^2 \\
y' = c^y_{0,0} + c^y_{1,0}x + c^y_{0,1}y + c^y_{1,1}xy + c^y_{2,0}x^2 + c^y_{0,2}y^2 
  \end{cases}
  \label{eq:polynomial_fit_2}
\end{align}

Given a polynomial order, the goal is to find the optimal set of weights $c^*$ that maximizes the similarity measure. This set of weights $c$ can be computed by two means: by directly looking into the space of parameters $c$ or by generating intermediate matching points that are used to perform a polynomial regression. 

While the first approach may be more direct, the manifold of parameters $c$ is very difficult to explore as a small change of parameters can lead to drastically different distortions. Therefore, in the proposed method, a set of matching points is generated, that is used to regress the polynomial coefficients $c$. As a small change in the location of the matching points lead to a close distortion, this process allows a better granularity in exploring the space of the distortions. More precisely, a regular grid and a distorted one are generated, and then the coefficients $c$ are determined by a least-squares fitting method that maps the distorted grid to the regular one. This function is described below:
\begin{itemize}
    \item Generate a distorted mesh grid $M = ( (x_0, y_0), \ldots, (x_N, y_N) )$
    \item Generate a regular mesh grid $M' = ( (x_0', y_0'), \ldots, (x_N', y_N,) )$
    \item Find the distortion $f$ parametrized by $c$ such as $c = \argmin_{\hat{c}} \sum_i^N ||\bigl(\begin{smallmatrix}
x_i'\\y_i'
\end{smallmatrix} \bigr) - f_{\hat{c}}(x_i, y_i) ||_2  $
\end{itemize}

where, $||.||$ is the euclidean norm. Note that the dimension of the mesh grid is tuned by the user via the number of points $N$ and/or the step-size between points. The influence of that parameter is discussed in section \ref{sec:discussion}. As a result, finding the optimal distortion $f^*$ requires to find the optimal set of matching points $M^*$ that better describes the underlying distortion. These matching points are then used to estimate the distortion itself. Counter-intuitively, the distorted grid may not match physical reality as discussed further in section ~\ref{subsec:repeatability}.

Finally, the \textit{polynomial distortion} function is called from the python package \textit{Sklearn}~\cite{sklearn} and \textit{Skimage}~\cite{skimage}. The next section describes how the pairs of matching points are created, using the CMA-ES optimizer.

\subsection{Using CMA-ES as a randomized black-box optimization to correct the distortions}
The goal of the optimization is to find the function $f$ that maximizes the $Similarity$ function between the speckles by finding the best distorted mesh grid $M$ (which is non-differentiable). To do so, a black-box optimizer method is used. Black-box optimization algorithms are efficient at solving non-linear, non-convex optimization problems, in continuous domains. They also overcome the deficiencies of the derivative-based methods in complex multidimensional landscapes that are rugged, noisy, have outliers or local optima, or when no error-gradient is available. All those features make the black-box optimizer relevant for the present problem, where the shape of the distortion is unknown and the process of superposing speckles intrinsically implies the $Similarity $ function to reach local maxima. The influence of noise present on the EBSD speckle is discussed in section \ref{subsec:missingspeckle}.

The black-box algorithm that is used in the present case is CMA-ES, which stands for Covariance Matrix Adaptation Evolutionary Strategy. CMA-ES has become a standard tool for continuous optimization and has been applied in various fields of research, such as image recognition for biology \cite{Ibanez2009,Sisniega2017}, energy \cite{Reddy2013}, chemistry \cite{Fateen2012,Weber2015}. Though, to the authors knowledge, it has not been applied in the field of electron microscopy. CMA-ES belongs to the family of evolutionary strategies (ESs), they are iterative algorithms, based on the principles of natural selection. CMA-ES involves a parametrized distribution (a multivariate normal distribution) that evolves throughout the iterations~\cite{Hansen1996}. 
ESs follow four distinct steps: initialization, sampling, evaluation and update. 
The initialization step consists in initializing the probability distribution parameters and sampling an initial population of individuals accordingly. At each iteration, a new \textit{generation} of individuals is created. Those individuals are candidate solutions to the optimization problem, which goodness can be evaluated on a \textit{fitness function}. After evaluation of the fitness of each individual in that population, the statistics of the distribution are updated according to the algorithm at hand. A new population is re-sampled according the new distribution and the process is repeated until a termination criterion is reached. Those operations are described below:\newline
\indent Initialize the distribution parameters $\theta$\\
\indent For each generation $g=0,1,...number\ of \ iterations$:\\
\indent \indent Sample $\lambda$ individuals $x_{1}, x_{2}, ..., x_{\lambda} \in \mathbb{R}^{n}$ according the probability distribution $P_{\theta}(x)$\\ 
\indent \indent Evaluate them on the fitness function\\
\indent \indent Update $\theta \leftarrow F(\theta, x_{1}, x_{2}, ..., x_{\lambda}, Similarity(x_{1}), Similarity(x_{2}), ..., Similarity(x_{\lambda}))$\\
where $P_{\theta}$ is a probability distribution that describes where the good solutions are believed to be and $F(.)$ is some update rule. In the CMA-ES, $P_{\theta}$ is a multivariate normal distribution \cite{Hansen1996}. Those steps are detailed in the following.

\paragraph{Initialization} The distribution $P_{\theta}$  encodes the distribution of the distorted spatial mesh grid. More precisely, the distribution is a multi-variate normal distribution where each dimension encodes either the x-coordinate or y-coordinate of a given point of the grid. For example, if the mesh is a 4x4 grid, the multi-variate normal distribution has 32 dimensions (16 points of 2 coordinates each). The initial distribution is defined by centering the mesh distribution on a regular grid. The standard deviation then encodes the initial acceptable moving distance of the matching points. Both the spacing of the points of the grid and the standard deviation are chosen by the user. In practise, CMA-ES slowly and iteratively distorts the regular grid, in order to retrieve a set of control (matching) points that encode the distortion. The concept of mesh grid here is to understand as an array of points that do not have any topological connectivity. In other words, the permutation of two points of the grid is allowed and does not affect the regression of the distortion function. The properties of the mesh are discussed in more detail in section \ref{subsec:repeatability}.

\paragraph{Sampling} New distorted meshes are generated following the distribution $P_{\theta}$. The coordinates are rounded, and matching points outside the range of the speckle dimension are kept. 

\paragraph{Evaluation} The fitness function is the $Similarity$ described in eq. \ref{eq:similarity}. The coefficients $c$ of the distortion function are first regressed by matching the distorted mesh grid to the regular one. The similarity measure is then computed and returned as the fitness score.

\paragraph{Update}
The basic CMA equation for sampling the search points at the generation $g+1$ is given in eq. \ref{eq:gaussian_mutation}.
\begin{align}
x_{k}^{(g+1)} \leftarrow m^{(g)} + \sigma^{(g)} \mathcal{N}(0, \mathcal{C}^{(g)}),\ for\ k=1, 2, ..., \lambda.
  \label{eq:gaussian_mutation}
\end{align}
where $x_{k}^{(g+1)} \in \mathbb{R}^{n}$ is the $k-th$ offspring of the generation $g+1$. $m^{g} \in \mathbb{R}^{n}$ is the mean value of the distribution at the generation $g$. $\sigma^{(g)} \in \mathbb{R}^{+}$ is the step size at the generation $g$. $\mathcal{N}(0, \mathcal{C}^{(g)})$ is a multivariate normal distribution with a mean of zero and covariance matrix $\mathcal{C}^{g}$. $\mathcal{C}^{g} \in \mathbb{R}^{n x n}$ is the covariance matrix of the distribution at the generation $g$. It is symmetric definite positive and describes the geometrical shape of the distribution. The initial value of $\sigma$, $\sigma_{0}$, is picked by the user and $\mathcal{C}_{0} = \mathcal{I}$. 
They both evolve throughout the iterations, as the population evolves. The self-adaptation of those parameters is the key point for the rapid convergence of the optimization\cite{Hansen2001}. The equations governing the update of $\sigma^{(g)}$, $\mathcal{C}^{(g)}$ and $m^{g}$ are described in \ref{App:CMA_update}. 
In the present case, this update shifts the spatial distribution of the points of the distorted grid in order to search for the optimal distortion function. 

\paragraph{Termination} This process is repeated until the termination criterion is reached. In the present case, a criterion based on the maximum number of steps is used. More advanced methods can be used to detect when the optimization (and thus the fitness score) starts plateauing. If several speckles are known to have the same distortion, they also can be used as a validation criterion to avoid over-fitting. Once the CMA-ES algorithm is finished, the means of the distribution $P_{\theta}$ are used as the final distorted mesh grid.



\subsection{Generation of the new EBSD map}
Once the maximum number of calls has been reached, the polynomial function leading to the best superposition of the speckles is applied to the EBSD data. In this process, the original grid of the EBSD is kept and the Euler angles, confidence indexes and other data is interpolated using the polynomial function. The points of the grid that end up containing no data are filled with zero-values for the confidence indexes and the $(0,0,0)$ triplet of Euler angles. The rotated and translated segmented BSE image is used to fill-in the phase data of the new EBSD file. For that reason, and as already mentioned, the segmentation of the BSE image has to be done as accurately as possible. Some application examples are shown in section \ref{sec:results}. 

\subsection{Source code and data availability}
The full distortion pipeline was developed in Python. The code source, the hyperparameters and one data set are available at \url{https://github.com/MLmicroscopy}.

\section{Results}
\label{sec:results}
All materials have been characterized in a FEI Versa 3D SEM, equipped with a TSL EDAX EBSD Hikari Plus Camera with an indexing speed of 300 to 400 frames per second and a 4x4 binning. The computer for the distortion correction was equipped with a Quad Core processor, 4.2 GHz and 64 GB of RAM. Most computations required about 10 to 15 minutes.

\subsection{$\beta$ phase in $\alpha - \beta$ Titanium alloy}
\label{Ti64}
A sample of Ti-6Al-4V with an $\alpha - \beta$ structure has been embedded in bakelite and electro-chemically polished. It was then mounted on the SEM stage and copper tape was used to enable electrical conductivity from the sample holder to the polished surface. An area of 14 x 20 $\mu m$ has been characterized with a 20 kV acceleration voltage and a step size of 40 nm. One purpose of collecting this dataset is to better resolve the phase map, using the better quality of the segmentation of the BSE image, with fine details. This mounting system was purposely used in order to induce a large drift of the beam during the EBSD scan.\newline

\begin{figure}[h]
\centering
\includegraphics[width=0.65\textwidth]{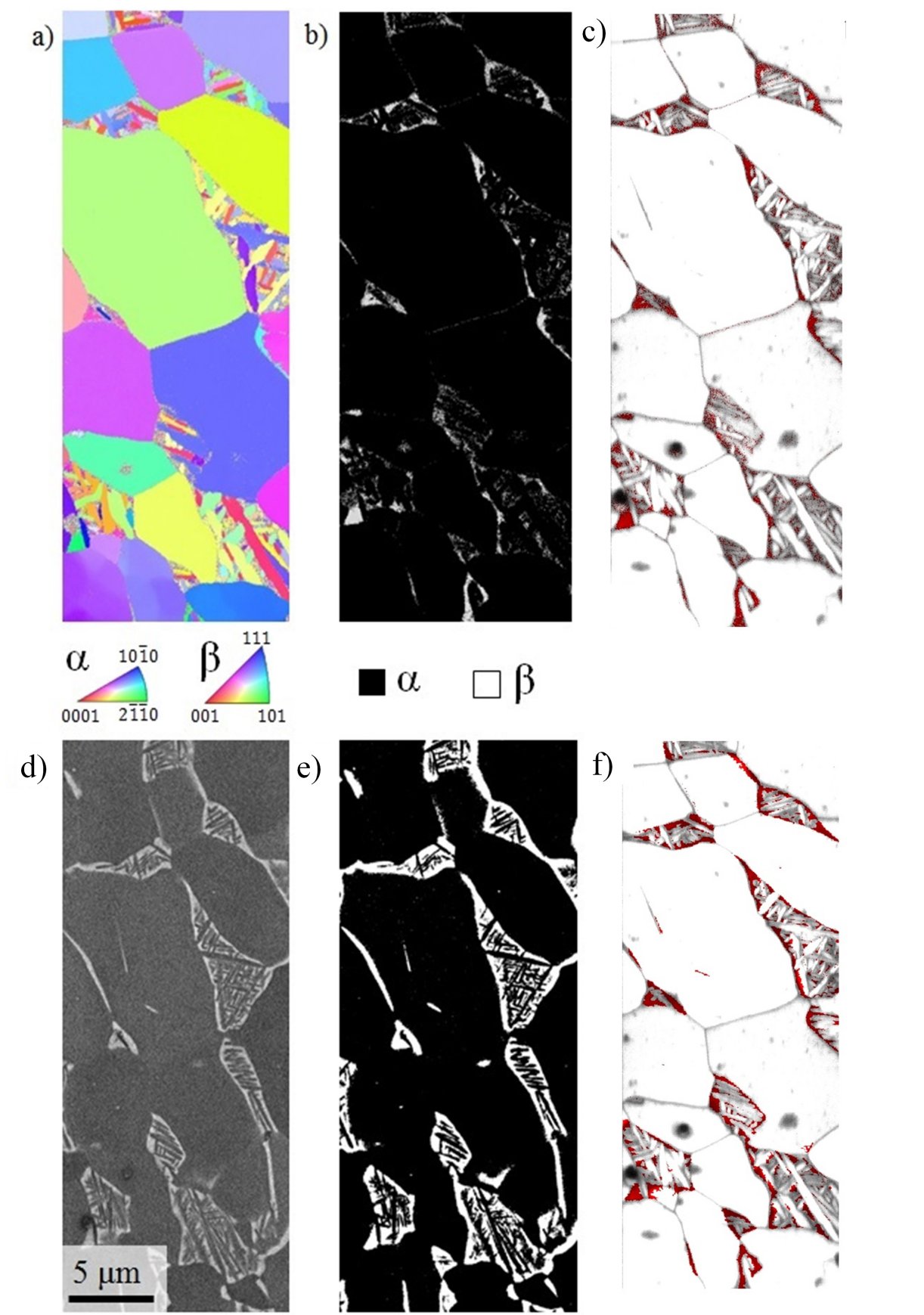}
\caption{Microstructure of the Ti-64 sample: a) Inverse Pole Figure (projected along the normal to the sample surface), b) EBSD speckle segmented out of the initial phase map, with $\alpha$ phase in black and $\beta$ phase in white, c) Initial phase map superimposed to the Image Quality map, d) BSE image used for phase segmentation, e) Segmented BSE image used as a phase map in the un-distorted EBSD file. f) Final phase map colored as c).}
\label{fig:Ti64all}
\end{figure}

The orientation map for this sample is shown on fig. \ref{fig:Ti64all}-a. The corresponding BSE image is shown on fig. \ref{fig:Ti64all}-d, where the $\beta$ phase exhibits a lighter contrast than the $\alpha$ phase. Fig. \ref{fig:Ti64all}-b shows the initial phase map with the $\beta$ phase in white. This phase map is superimposed to the Index Quality map on fig. \ref{fig:Ti64all}-c, where the $\beta$ phase is colored in red. The comparison of fig. \ref{fig:Ti64all}-b and -e shows that only a small fraction of the $\beta$ phase was indeed identified as such on the initial EBSD data. Its initial area fraction is 3.4\%, versus 9.9\% in the segmented BSE image. The CMA-ES optimizer has been applied on a 25 x 25 points mesh grid, with an initial standard deviation $\sigma_{0}=20$ pixels, using a polynomial order of 3 and 5,000 iterations. Fig. \ref{fig:Ti64all}-f shows the final phase map, colored as fig. \ref{fig:Ti64all}-c. It superimposes well with the Index Quality map. The final area fraction of $\beta$ phase is 9.9\%, as in the BSE speckle. The contribution of the CMA-ES optimizer to compensate higher order distortions and match all the features of the speckles is discussed in section \ref{subsec:versatility}.

\subsection{$\gamma$ and $\gamma '$ phases in Rene 65 superalloy}
\label{subsec:R65}
\begin{figure}[!htb]
\centering
\includegraphics[width=0.85\textwidth]{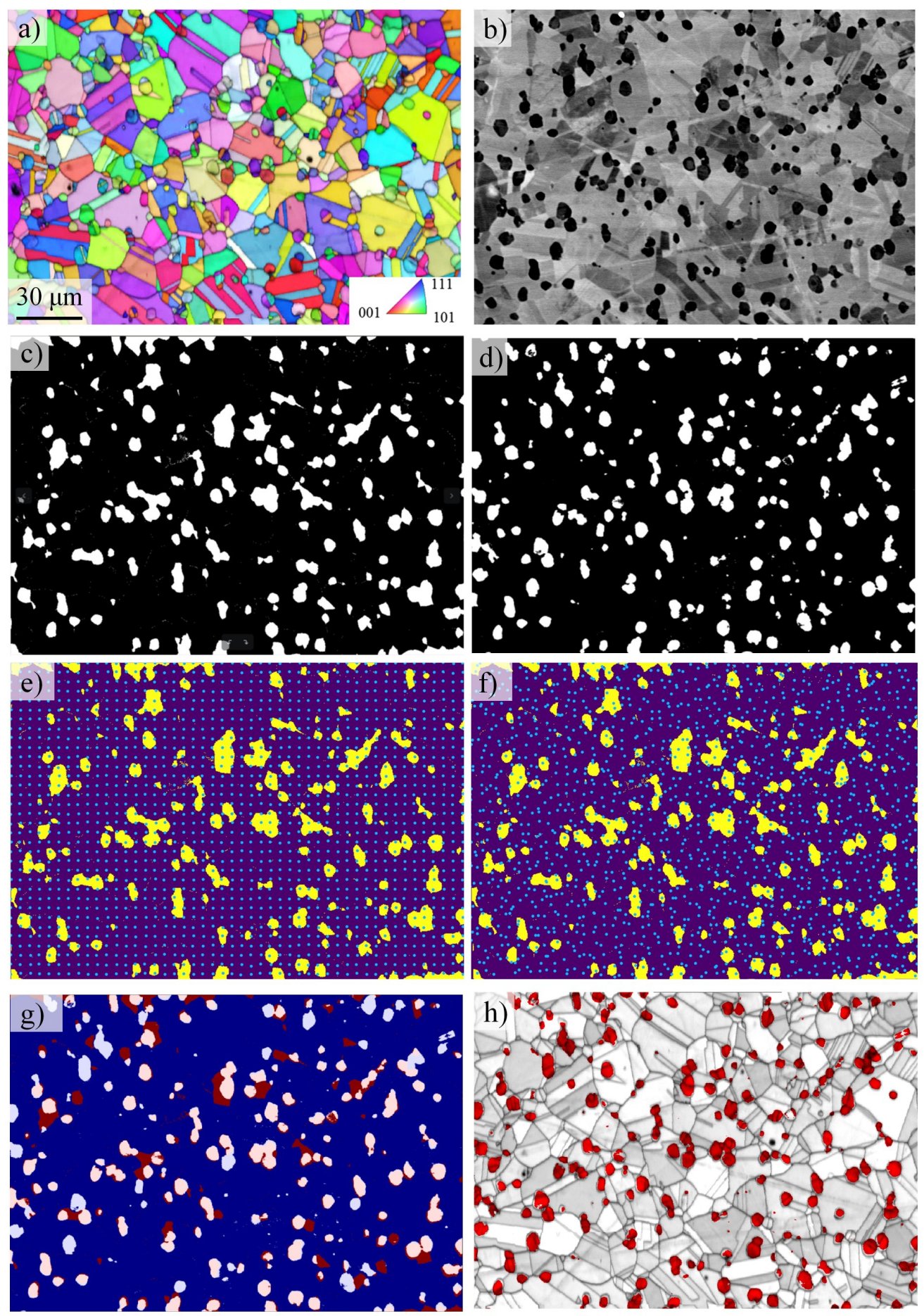}
\caption{Microstructure of the solution annealed Rene 65 sample. a) EBSD map colored according to the Inverse pole figure projected along the normal direction to the surface of the sample, b) BSE image from which the precipitates have been segmented, c) BSE speckle, d) EBSD speckle: smallest features of the dataset segmented from the initial EBSD file, considering annealing twins as internal defects, e) initial mesh over the EBSD speckle, f) final mesh over the EBSD speckle, g) Superimposed EBSD (blue) and BSE (red) speckles, h) Phase map with the precipitates colored in red superimposed to the pattern quality map of the new EBSD file.}
\label{fig:R65all}
\end{figure}

Rene 65 is a polycrystalline Nickel-based alloy that has been designed for turbine disk applications. Its chemical composition (wt\%) is Ni-16 Cr – 13 Co – 3.7 Ti – 2.1 Al – 4 Mo – 4 W – 1 Fe – 0.7 Nb – 0.05 Zr – 0.016 B \cite{Heaney2014}. The sample was extracted from a ring that was subjected to a sequence of forging operations followed by solution annealing and aging treatments, 1065$\degree$C for one hour and 760$\degree$C for 8 hours respectively. It was prepared using conventional metallography techniques, followed by vibratory polishing using a 0.04 $\mu m$ $Al_{2}O_{3}$ suspension. Its microstructure consists of fine $\gamma$ matrix grains, which average equivalent diameter is 10-12 $\mu m$. Spherical primary $\gamma '$ precipitates are located on the grain boundaries, with an equivalent diameter in the range of 1-4 $\mu m$. EBSD maps and corresponding BSE images have been acquired using an acceleration voltage of 20 kV with a step size of 0.1 $\mu m$ over an area of 150 x 200 $\mu m$. Figure \ref{fig:R65all}-a shows the initial EBSD map colored according to the orientation of the crystals projected along the normal to the polished surface. Figure \ref{fig:R65all}-b shows the corresponding BSE image on which the precipitates exhibit a darker contrast. Figure \ref{fig:R65all}-c and -d show the BSE and EBSD speckles, respectively, created by segmenting the $\gamma '$ precipitates. Fig. \ref{fig:R65all}-c has been created by segmenting  the features which equivalent diameter is smaller than 5 $\mu m$, considering the annealing twin boundaries as internal defects. This criterion enables a rough segmentation of the precipitates. Note that some precipitates may have been omitted from the segmentation and some small grains may have been mistakenly segmented as precipitates. The influence of the quality of the EBSD speckle on the goodness of the compensation of the distortions is discussed in section \ref{subsec:missingspeckle}. This segmented image was then used as the EBSD speckle for the CMA-ES optimizer, along with a segmented BSE image of the same area. The CMA-ES optimizer was used on a mesh grid with a spacing of 24 pixels between consecutive points, an initial standard deviation of 5 pixels and a polynomial order of 3. The initial grid is presented on figure \ref{fig:R65all}-e, the final grid is presented on figure \ref{fig:R65all}-f. The superimposed speckles after applying the CMA-ES optimizer are shown on fig. \ref{fig:R65all}-g. The distortion correction leads to a final score of 0.62 after 1,800 iterations. Figure \ref{fig:R65all}-h shows the final phase map with the precipitates colored in red, superimposed with the Image Quality map displayed in grey scale. The location of the precipitates corresponds well with the location of the grain boundaries on the Image Quality map, over most of the area of interest. The precision of the method is discussed in sections \ref{subsec:gamma_gammaprime} and \ref{subsec:repeatability}.

\subsection{Pores in Additive Manufactured Inconel 718}
\label{subsec:AM718}
\begin{figure}[!htb]
\centering
\includegraphics[width=0.6\textwidth]{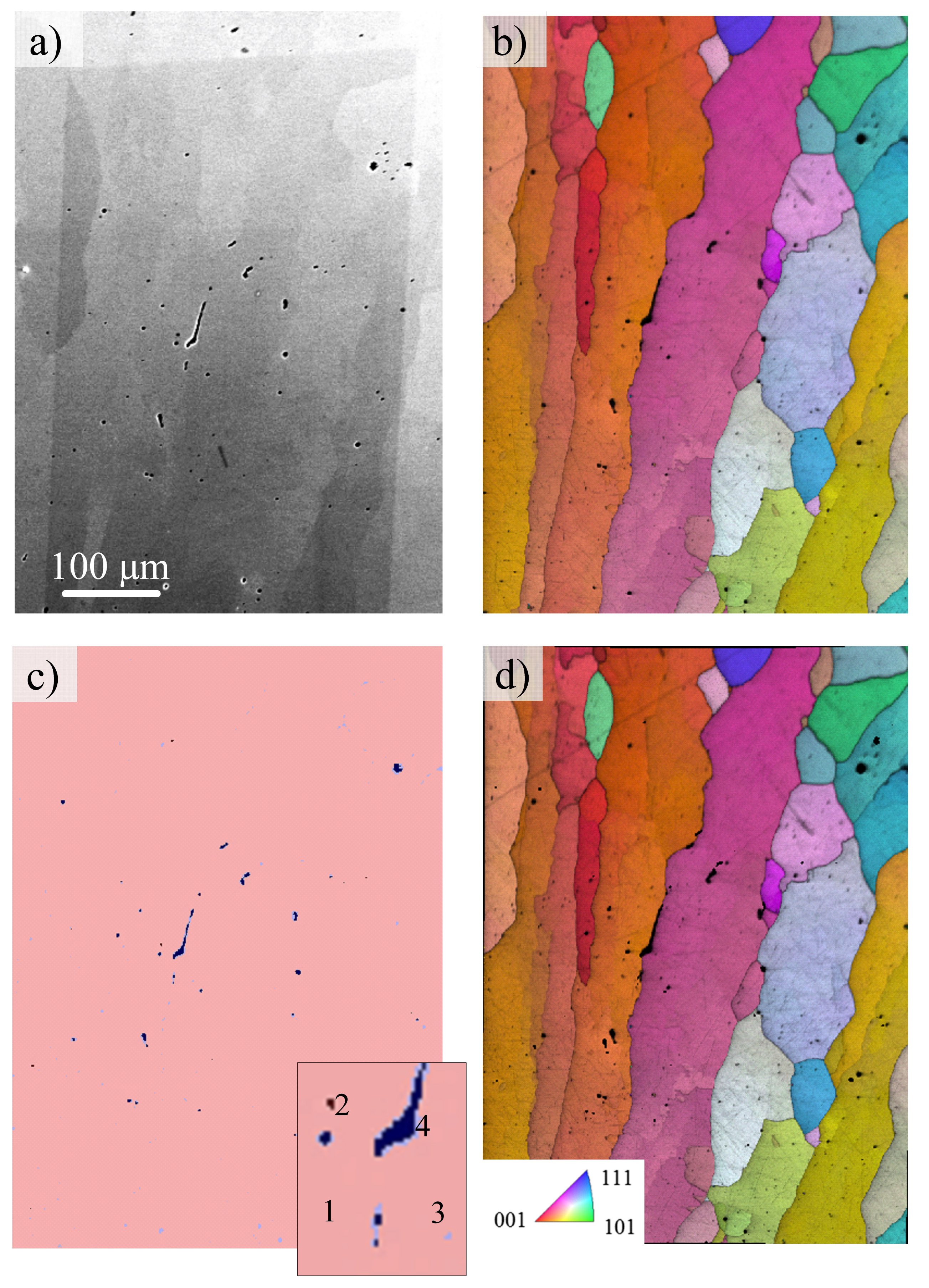}
\caption{Microstructure of Additive Manufactured Inconel 718, a) SE image of used for the segmentation of the pores, the trapezoid contamination zone corresponding to the EBSD area is visible, b) Initial EBSD map colored according to the Inverse Pole Figure projected along the build direction superimposed to the Image Quality map in greyscale, c) Superimposed SE (red) and EBSD (blue) speckles after CMA-ES, d) Final EBSD map superimposed to the Image quality map in greyscale, with pores colored in black.}
\label{fig:AM718}
\end{figure}

The alloy Inconel 718 has been manufactured by the Electron Beam Melting (EBM) technique, as described in \cite{Kirka2017}. A sample was cut by Electrical Discharge Machining (EDM) and polished parallel the build direction, using conventional metallography processes followed by 0.05 $\mu m$ colloidal silica. Image and EBSD data have been acquired under a 20 kV acceleration voltage, using a 0.5 $\mu m$ step size over an area of 300 x 700 $\mu m$. Fig. \ref{fig:AM718}-a shows the Secondary Electrons (SE) image that was used for the SE speckle. The beam contamination zone corresponding to the area of the EBSD map exhibits a trapezoid shape. A standard cleaning procedure involving slight cleaning of the EBSD data has been applied, consisting in removing isolated groups of less than 3 pixels having the same orientation, leading to a change of 1\% of the data. Since the pores could not be indexed, the TSL software assigned random orientations to those pixels. The inverse pole figure map, projected along the build direction, is shown on fig. \ref{fig:AM718}-b. Some of the grains have their $<001>$ axis preferentially aligned with the build direction (vertical), a common feature of additive manufactured face-centered cubic materials. A partial EBSD speckle of the pores has been created by segmenting the pixels of lowest Confidence Indexes (CI). On the other hand, the pores have been accurately segmented from a SE image of the same area, in order to create the SE speckle. The CMA-ES strategy has been applied, using a 65 x 65 points mesh grid, an initial standard deviation $\sigma_{0} = 20$ pixels, a maximum sampling of 5,000 iterations and a polynomial order of 3 for the distortion function. Fig. \ref{fig:AM718}-c shows the superimposed speckles after the CMA-ES strategy. The EBSD speckle consists of light blue points on a transparent background, and the SE speckle consists of dark red points on a light red background. Thus, as shown in the insert on the figure: light red pixels (labeled "1" on the insert) correspond to the SE background, dark red pixels (labeled "2") correspond to pores of the SE speckle that did not match any pores of the EBSD speckle, light blue pixels (labeled "3") are pores of the EBSD speckle that did not match the SE speckle dark blue pixels (labeled "4") are pores that were successfully matched. \footnote{using a method based on the CI to segment the pores can lead to inaccuracies in the segmentation: some pixels may have been segmented as pores even though they are not (typically those pixels correspond to grain boundaries).} From this basis, the final EBSD map is shown on fig. \ref{fig:AM718}-d. Only the points with a confidence index higher than 0.2 are plotted. The color code is the same as fig. \ref{fig:AM718}-b and the pores are colored in black. Most of those pores are located on the grain and sub-grain boundaries. After applying the distortion, some data points ended up out of the initial grid, which leads to the black pixels on the edges of fig. \ref{fig:AM718}-d. This dataset illustrates the limits of this method to compensate the distortions over broad areas when the features are unevenly distributed over the area.

\section{Discussion}
\label{sec:discussion}

\subsection{Versatility of the method}
\label{subsec:versatility}

\begin{figure}[h]
\centering
\includegraphics[width=0.7\textwidth]{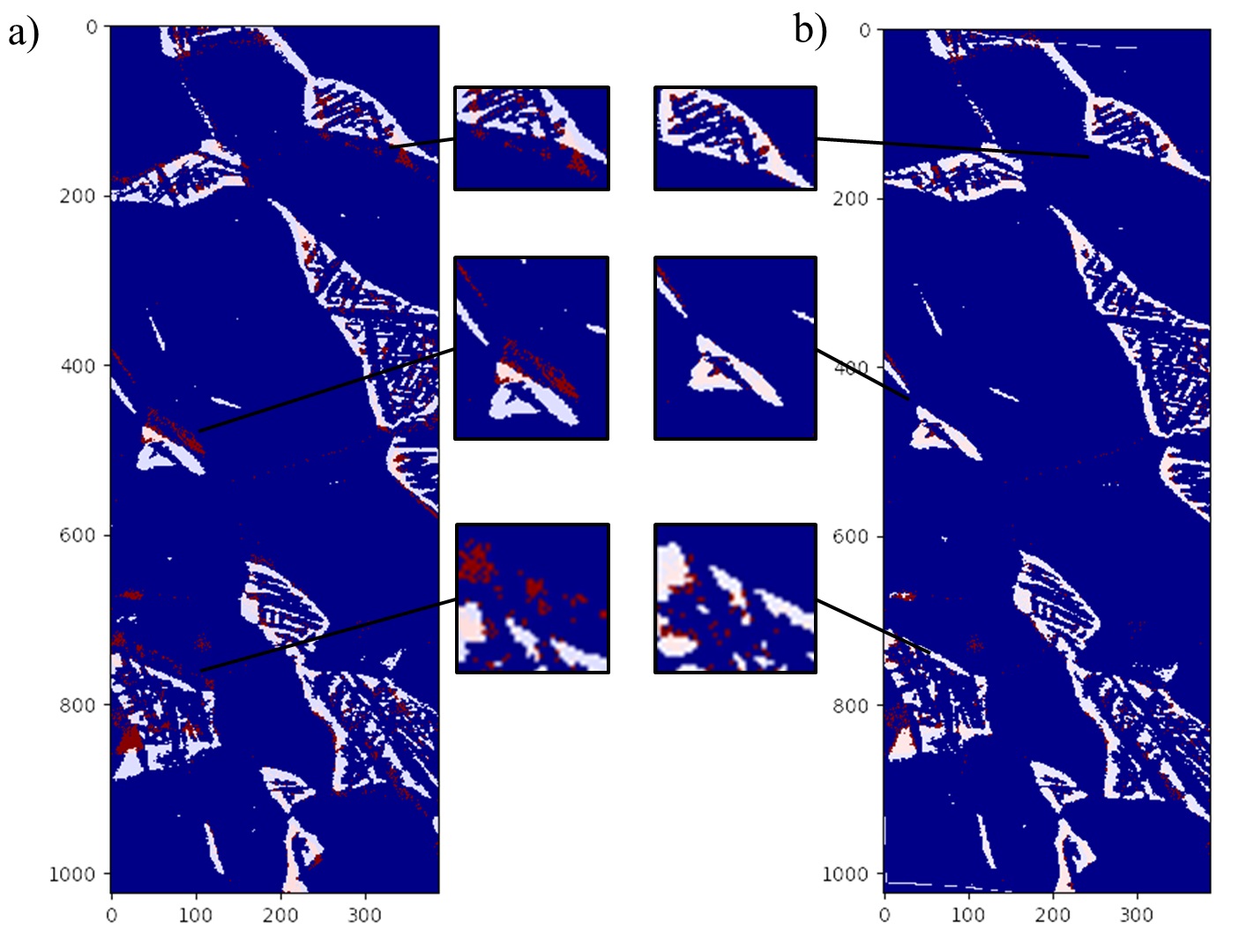}
\caption{Speckle superposition after correction of a) Affine distortions only, b) All distortions. On a blue background, the EBSD speckle is colored in red and the BSE speckle in white.}
\label{fig:Ti64comparison}
\end{figure}

Several methods have been proposed in the literature, for correcting the distortions in EBSD data, using various algorithms to make up for the drift distortions. Zhang et al. \cite{Zhang2014} have shown that a thin plate spline function enables compensation of the distortions in EBSD data. Contrary to most methods, the use of the CMA-ES strategy does not assume any specific shape of the distortion function, nor constrain the order of the polynomial function. The use of speckles and a score to quantify the goodness of the superposition enables the matching of both images at the resolution of the pixel. Fig. \ref{fig:Ti64comparison} shows a comparison between the superposition of the speckles in the Ti6Al4V dataset, after the compensation of affine distortions only (a) and after the CMA-ES procedure (b). On a blue background, the EBSD speckle is colored in red and the BSE speckle in white. Compensating the affine distortions enables a good overlap of the speckles. The CMA-ES optimizer then compensates the finer distortions, even very local, and enables to achieve a better matching of the speckles despite an important drift of the beam. Three examples are presented in the inserts of Fig. \ref{fig:Ti64comparison}, where the laths are accurately superimposed.\newline

\begin{figure}[h]
\centering
\includegraphics[width=0.25\textwidth]{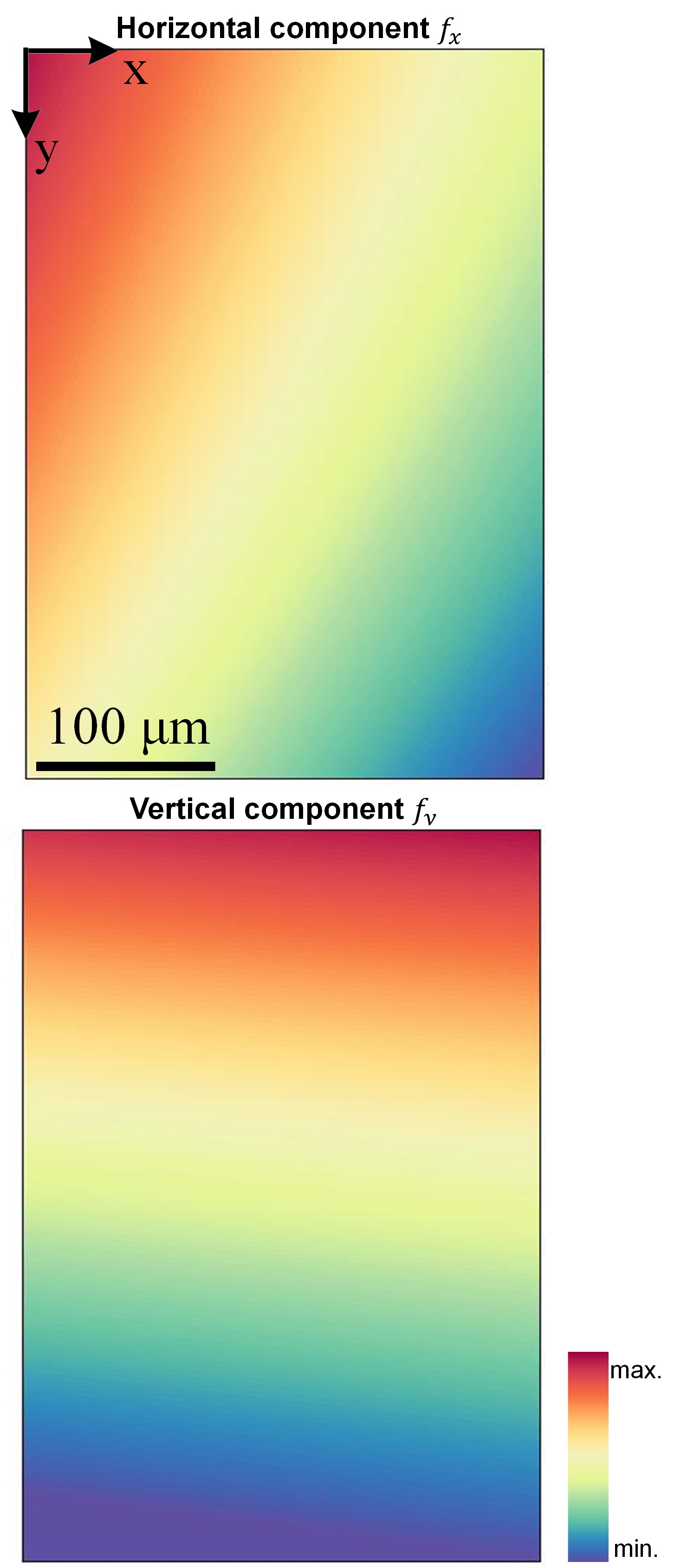}
\caption{X and Y components of the distortion function in the Additive Manufactured Inconel 718 sample.}
\label{fig:XandYcomponents}
\end{figure}

Figure \ref{fig:XandYcomponents} shows the maps of the vertical and horizontal components of the distortion function in the Inconel 718 material. The origin and x and y directions of the scan are indicated with arrows. The electron beam scanning pattern starts from the top left corner and goes line-by-line from top to bottom. A polynomial order of 3 has been used. Both maps show that the maximum error occurs on the first points of the scan. Indeed, drift phenomena are the highest in the beginning of the scan. The magnitude of both components then decrease along both x and y directions. Along the x direction, $f_{x}$ decreases relatively more than $f_{y}$, which means that the error on the x' coordinate decreases along this direction, whereas the error on the y' coordinate remains relatively constant. From a line to another (along the y direction), the magnitude of $f_{x}$ decreases, which means that the error on the location of the x' coordinate decreases from a line to another. A similar observation can be made about the vertical component $f_{y}$. Those trends are consistent with most drift phenomena, where the first scanned pixels that are the most subjected to drift. Drift phenomena tend to decrease as the scan proceeds: along a line, and from a line to another. A quantitative analysis of those trends can be found in \ref{App:dFxdFy}.

\subsection{Sensitivity to the choice of the initial parameters in CMA-ES}
\label{subsec:meshrefinement}

\begin{figure}[h]
\centering
\includegraphics[width=0.60\textwidth]{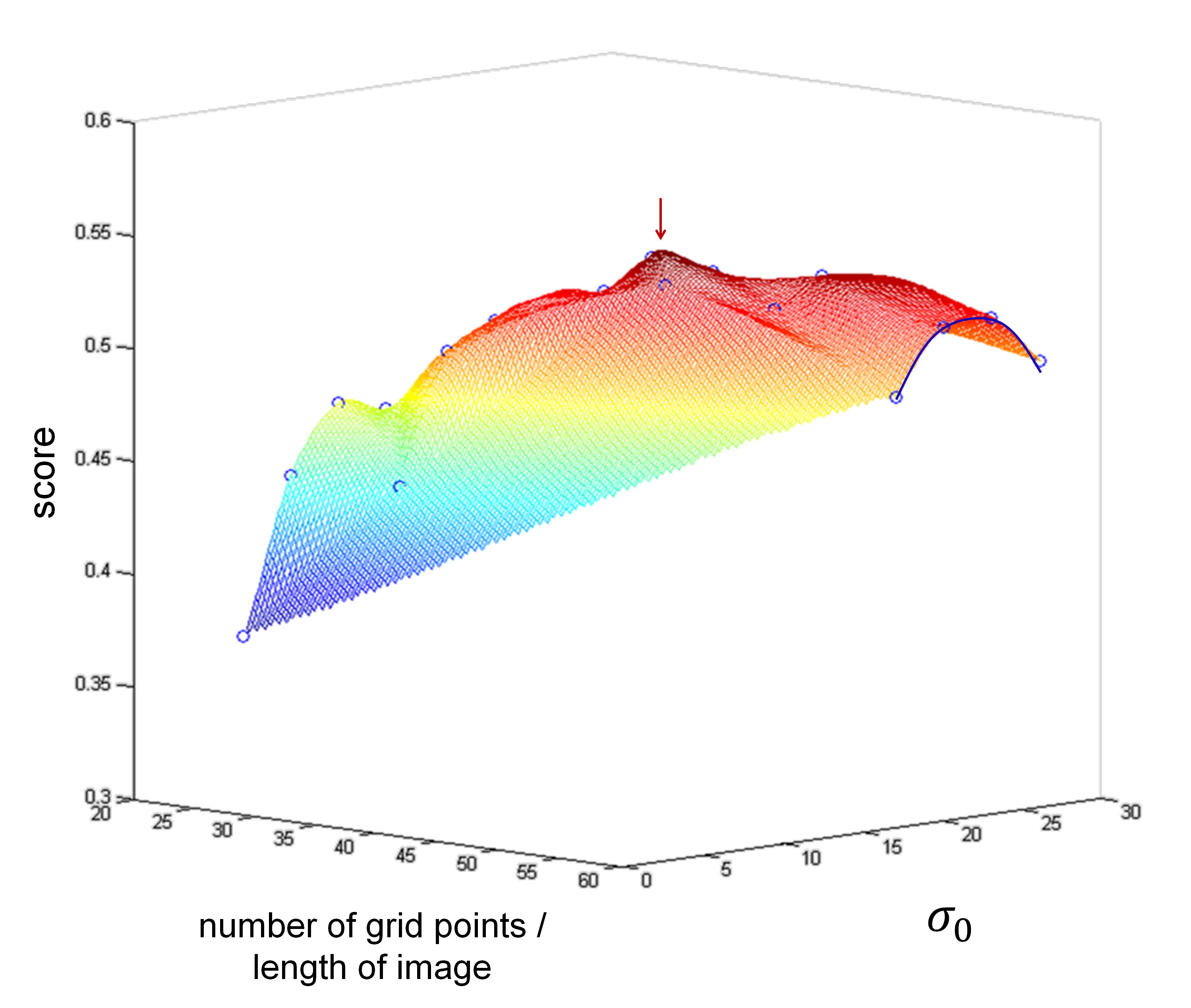}
\caption{Influence of the initial CMA-ES parameters on the final score, in the Ti-6Al-4V dataset.}
\label{fig:meshSensitivity}
\end{figure}

The influence of the CMA-ES parameters on the goodness of the matching has been studied in the Ti-6Al-4V dataset. Figure \ref{fig:meshSensitivity} shows the final scores obtained for different numbers of points per width and height of the image, and initial step sizes. The obtained surface is convex, with a global maximum reached for (35,21), pointed by the red arrow on the figure. For a fixed number of grid points, the score admits a maximum for a given initial step size $\sigma_{0}$. This behavior is visible on the four points on the far right of the figure, on the part of the envelope highlighted by a blue curve. For 55 grid points per length of the image, the score reaches a maximum of 0.5128 for an initial step size of 27 pixels. A similar behavior occurs for any given initial step size but varying numbers of points in the mesh grid. Note that a similar behavior is expected for all data sets, but for different values of the parameters.

\subsection{Precision of the reconstruction}

\begin{figure}[!htb]
\centering
\includegraphics[width=0.60\textwidth]{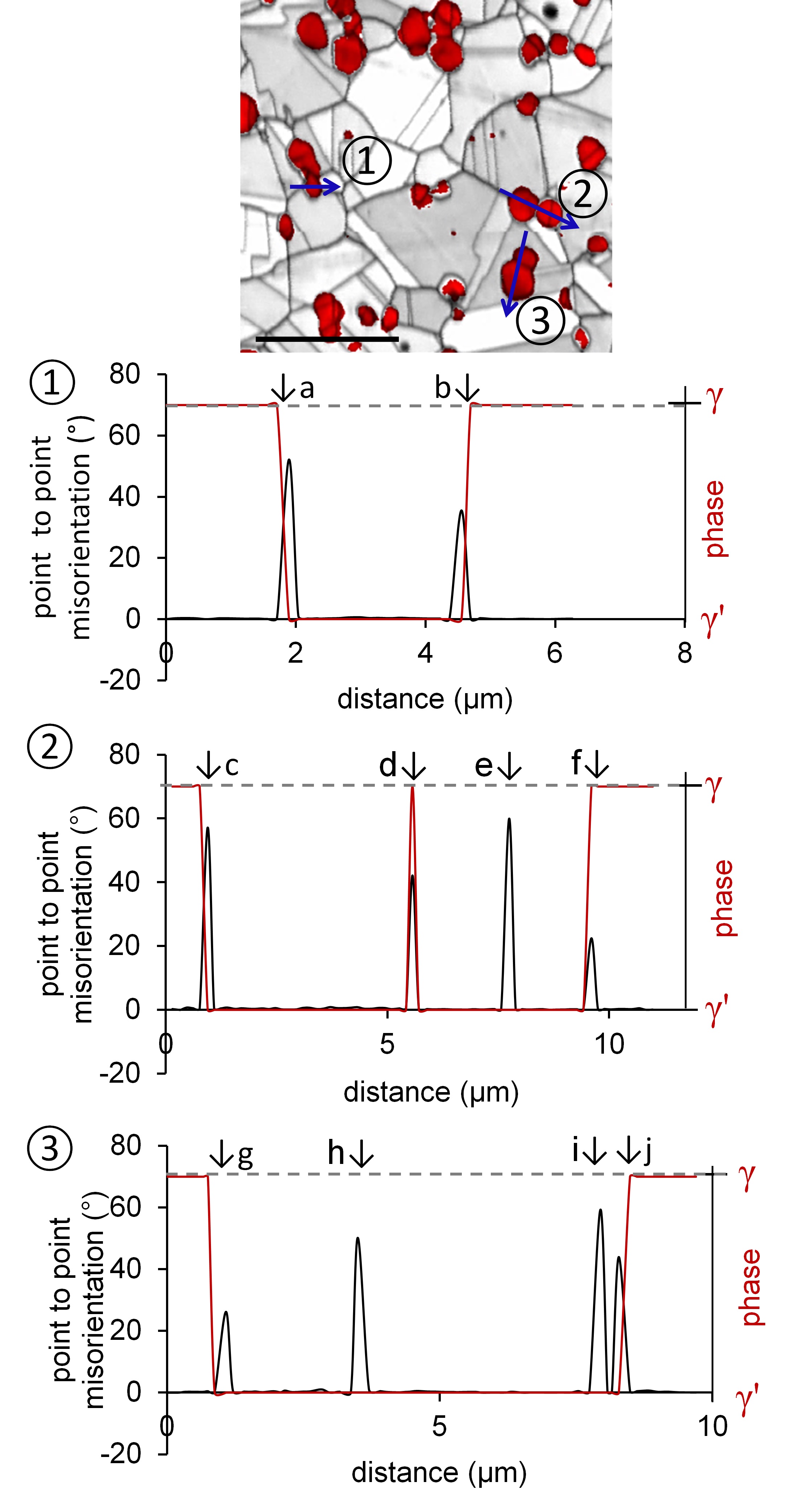}
\caption{Comparison of the location of phase and grain boundaries in the reconstructed EBSD data of the Rene 65 alloy, on three misorientation profiles across $\gamma '$ precipitates.}
\label{fig:R65misorientation}
\end{figure}

The accuracy of the reconstruction of the phases can be quantified by comparing the location of the phase boundaries and grain boundaries on the reconstructed maps. Such measurements have been performed on the Rene 65 dataset and displayed on fig. \ref{fig:R65misorientation}. A subset of the corrected EBSD map has been selected where three line profiles have been traced, crossing isolated or clusters of $\gamma '$ precipitates. The point-to-point misorientation is plotted in black and superimposed to the phase data, which is plotted in red. The first profile crosses a single precipitate. The phase boundaries and grain boundaries "a" and "b" are indicated on the corresponding profile. The transition from a phase to the other corresponds well with the crossing of high-angle boundaries. The second profile crosses a cluster of two precipitates which were separated on the segmentation of the BSE image. The phase boundaries "c", "d" and "f" are well superposed with the grain boundaries. Note that the boundary "d" consists of a single pixel. The profile also crosses an annealing twin boundary ($\Sigma 3$, 60$\degree$ rotation around a $<111>$ axis) present in one of the precipitates at the marker "e", characterized by a 60$\degree$ misorientation. The third profile also crosses two precipitates which were not separated on the segmented BSE image. The grain boundaries of the first precipitate are labeled "g" and "h" on the profile. The second precipitate contains an annealing twin boundary "i", that is crossed right before the grain boundary "j". This grain boundary is well superimposed with the phase boundary.
The error on the location of the phase boundaries has been measured on all the profiles, by comparing their location $x_{Ph.B}$ to that of the grain boundaries $x_{GB}$ through the parameter $\Delta x$ of eq. \ref{eq:deltaX}. The results are displayed in table \ref{tab:deltaX}. The error is less than 0.2 $\mu m$ for all the boundaries (i.e. 2 pixels), reaching $\approx$ 0 $\mu m$ for most of them. The error on the location of the two phase boundaries that were not exactly correlated with the grain boundaries ("f" and "g") corresponds to a misplacement of 1 and 2 pixels respectively. This gives an indication of the precision achievable with this correction method, assuming a good segmentation of the BSE image is achieved. A segmentation in which the boundaries of the objects have been eroded or blurred impacts the quality of the reconstruction.

\begin{align}
\Delta x = \|x_{Ph.B} - x_{GB}\|
\label{eq:deltaX}
\end{align}

\begin{table}
\caption{Precision of the location of the phase boundaries versus grain boundaries in the profiles of Fig. \ref{fig:R65misorientation}, in the Rene 65 dataset}
\begin{center}
\begin{tabular}{ |c|c|c| }
 \hline
 Boundary & $\Delta x$ ($\mu m)$\\ 
 \hline
 a & 0.0 \\ 
 b & 0.0 \\ 
 c & 0.0 \\
 d & 0.0 \\
 f & 0.1 \\
 g & 0.2 \\
 j & 0.0 \\
 \hline
\end{tabular}
\end{center}
\label{tab:deltaX}
\end{table}

\subsection{Convergence, stability and repeatability of the CMA-ES optimization}
\label{subsec:repeatability}

During the CMA-ES optimization procedure, new distorted mesh grids are generated to regress the polynomial distortion. Counter-intuitively, these grids are not constrained to encode a realistic distortion; they are only optimized such that the final distortion is meaningful. Thus, one must not use the distorted mesh grids outside the polynomial regression step, as several different mesh grids can lead to the same distortion function. Indeed, there exists an infinite number of sets of points that extrapolate to the same final function. Those points do not have to always be on the curve itself. Even though phenomena of grid points swapping does not affect the calculation of the distortion function, the authors suggest that the user sets the $\sigma_{0}$ parameter as less than the distance between two consecutive grid points, if the user wants to avoid this phenomenon. To a lesser extent, several polynomial coefficients may also lead to a very similar resulting distortion function, only differing around the border of the speckle. 

\begin{figure}[!htb]
\centering
\includegraphics[width=0.55\textwidth]{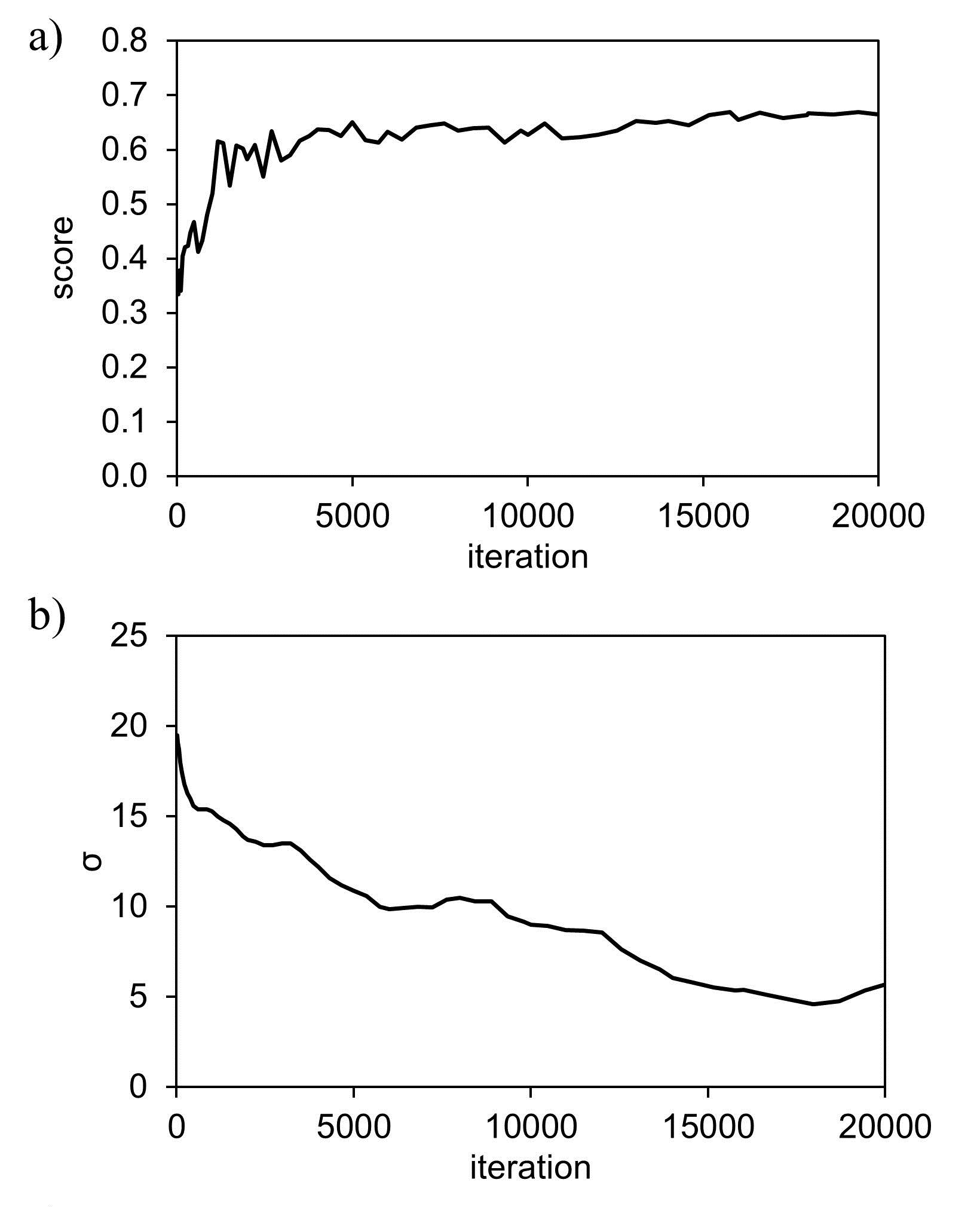}
\caption{Typical optimization path with CMA-ES, as a function of the iterations, here on the AM Inconel 718 dataset: a) Value of the similarity function at the generation $g$, b) $\sigma^{(g)}$: step size at the generation $g$.}
\label{fig:CMAESprocess}
\end{figure}

Despite this apparent limitation, CMA-ES turns out to be highly reproducible in practice. For example, fig. \ref{fig:CMAESprocess} shows the typical pattern of convergence of the CMA-ES algorithm: fig. \ref{fig:CMAESprocess}-a shows the evolution of the score as a function of the number of iterations in the CMA optimizer and \ref{fig:CMAESprocess}-b shows the evolution of the step size throughout the iterations. Starting from a score value of about 0.4, the CMA optimizer enables a quick convergence towards a steady-state value of about 0.7. This steady state is reached after less than 5000 iterations. As the score increases, the step-size $\sigma$ decreases with the number of iterations, indicating little variation around the mesh grid ground.

\begin{figure}[!htb]
\centering
\includegraphics[width=0.55\textwidth]{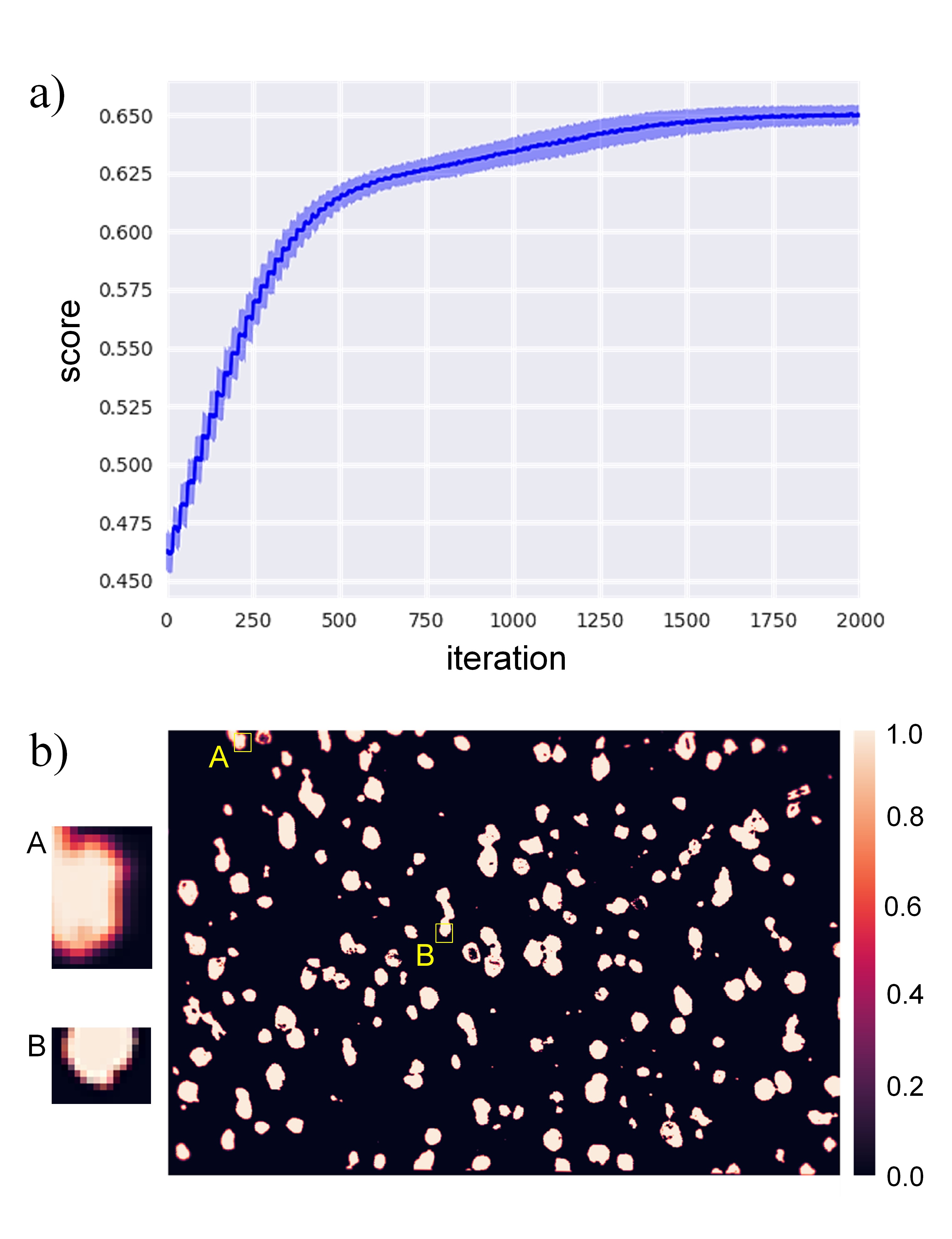}
\caption{Convergence and stability of the CMA optimization over 100 runs in the Rene 65 dataset. a) Mean score (dark blue), minimum and maximum values (light blue), as a function of the number of iterations,  b) Corresponding heatmap : the pixels are colored according to the number fraction of times they were identical among the 100 runs.}
\label{fig:repetition}
\end{figure}

The CMA-ES strategy has been applied 100 times on the Rene 65 dataset. Each run consisted in 2000 iterations, with a step size of 75 points, an initial standard-deviation $\sigma_{0}$ of 5 pixels. A polynomial order of 3 was used for the polynomial function. The CMA optimizer produced an average score of 0.6587 with an associated standard deviation of 0.0015. The similarity score between distorted speckles over the 100 runs was 0.9353 in average, indicating a good precision and repeatability of the process. Figure \ref{fig:repetition}-a shows the evolution of the score throughout the iterations: the dark blue curve corresponds to the mean score over 100 runs. The light blue colored area around the curve corresponds to the lower and upper bounds of the score (mean - standard deviation, and mean + standard deviation, respectively). Those values show that the optimization is stable and repeatable over several runs. Some steps are clearly visible on the first 500 iterations and correspond to the new bounds determined by the creation of new individuals at each generation in CMA-ES. Based on those statistics, a heat map has also been generated, displayed on fig. \ref{fig:repetition}-b. This map shows the segmented BSE speckle colored according to the number fraction of times that a pixel was present at a given (x', y') location. In other words, a pixel that appears in white (value 1.0) was always assigned to the $\gamma '$ phase. On the opposite, a pixel that appears in black was never assigned to this phase. The more consistent and precise the optimization, the sharper the contrast on this map. The shape of the precipitates appears clearly on fig. \ref{fig:repetition}-b, which is consistent with the good repeatability suggested by fig. \ref{fig:repetition}-a. This map also gives information about the precision of the reconstruction over the whole map. The center of the precipitates usually appears with a value of 1.0, however their boundaries are usually less precise. The deviation on the reconstruction of the boundaries of the precipitates is illustrated on two examples, labeled "A" and "B". The precipitate "A" is close to the border of the image, and the precipitate "B" is in the center. The inserts on the side of the figure show that the boundaries of those two precipitates are not reconstructed with the same consistency over the 100 slices. There is only one layer of pixels having a value lower than 1.0 on the precipitate B, versus 2 to 3 rows on the boundary of the precipitate A. The boundaries of most of the precipitates located on the edges of the map have a similar coloring. This indicates that the CMA-ES optimizer leads to consistent results in the center of the map, with a deviation of about 1 pixel as the location of the phase boundaries are approached. At the borders of the map, the consistency on the location of the phase map is about 2 to 3 pixels. This corresponds, in this dataset, to an associated error of 0.2 to 0.3 $\mu m$ on the location of the phase boundaries. This error remains much smaller than the actual size of the features of interest. In other words, the reconstruction does not artificially create additional features (precipitates) nor assign the wrong phase to any feature.

Again, it should also be noted that different final grids can describe the same polynomial distortion function, since the calculation of the function is not affected by the permutation of the points of the mesh grid. Moreover, different polynomial functions can describe the same physical distortion. This phenomenon is enhanced if the points of the speckle are sparsely distributed or mostly located in the center of the region of interest.

\subsection{Sensitivity to the accuracy of the EBSD speckle}
\label{subsec:missingspeckle}
\begin{figure}[!htb]
\centering
\includegraphics[width=0.45\textwidth]{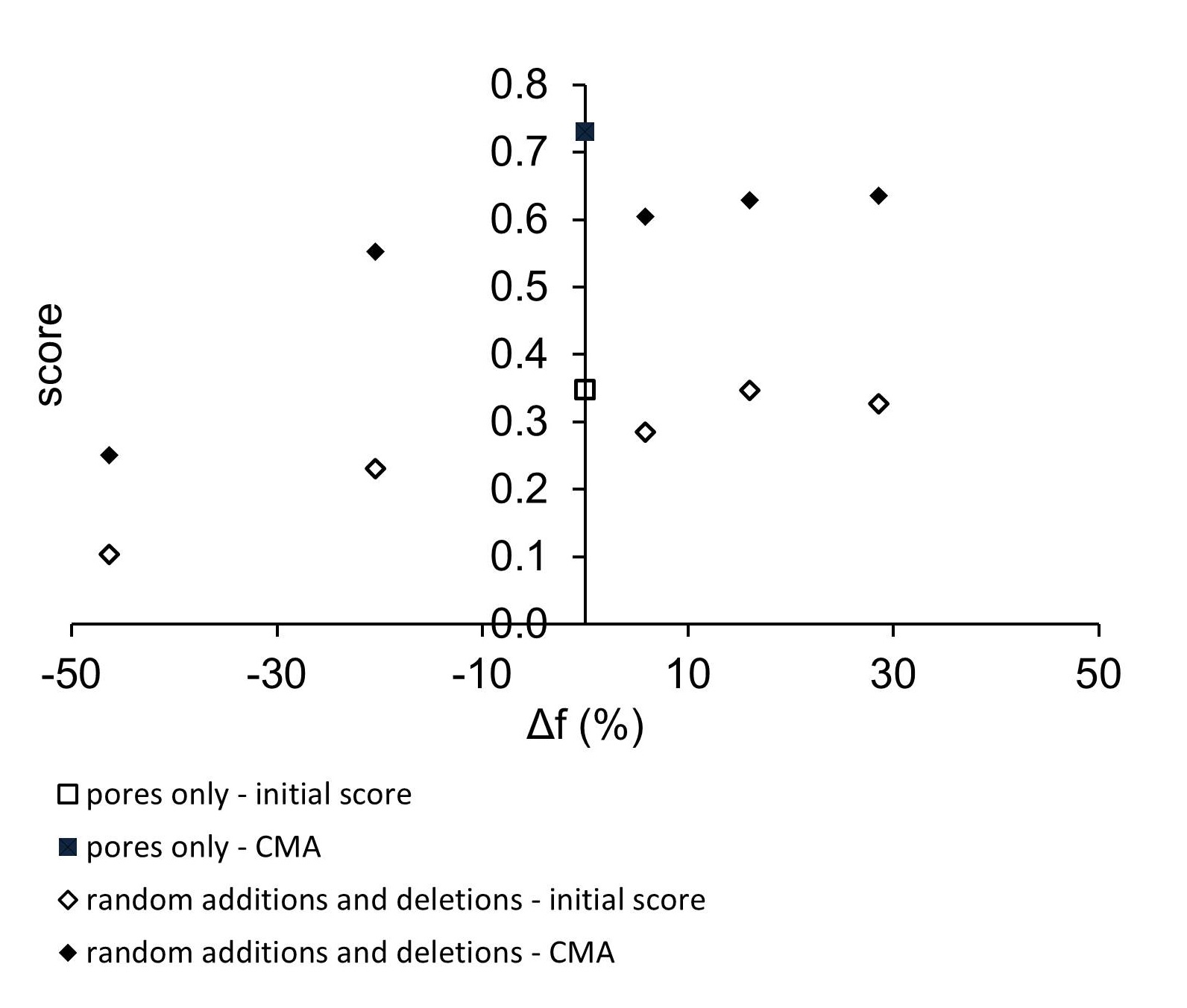}
\caption{Influence of the addition or deletion of points on the EBSD speckle. Initial scores and final scores in CMA-ES as a function of the percent of points added or deleted.}
\label{fig:pointdeletion}
\end{figure}

As mentioned previously, the use of speckles eliminates the manual selection of reference points. Another advantage of using speckles and a score based on the number of superimposed pixels, is that the EBSD speckle can be only a small subset of the original image. This is the case in the Rene 65 dataset, where the BSE speckle contains all the $\gamma'$ precipitates but the EBSD speckle contains the smallest features in the dataset. Some precipitates are missing on this speckle, and small grains are mistakenly segmented as precipitates. This does deteriorate the theoretical maximum score achievable, but does not prevent a good superposition of the speckles, as already shown on fig. \ref{fig:R65all}-c. The influence of the quality of the EBSD speckle has been studied on the Inconel 718 dataset: some noise was added (wrong features and random noise) and some pores were also removed from the EBSD speckle on purpose, in comparison with the one used on \ref{subsec:AM718} (fig. \ref{fig:AM718}). The amount of features added or deleted is quantified by the $\Delta f$ parameter, which quantifies the percentage of pixels points added or deleted to the speckle containing the correct number of pores: $\Delta f = \frac{number of pixels in wrong speckle}{number of pixels in right speckle}$. Figure \ref{fig:pointdeletion} shows the best score obtained as a function of the percentage of pixels added or deleted in the EBSD speckle. Empty and filled markers correspond to the initial and final scores in CMA-ES, respectively. The speckle at $\Delta f=0\%$ corresponds to the EBSD pattern in which only the pores were segmented. It leads to the best score achievable. The deletion of points leads to a drastic decrease of the score. The addition of "wrong" cavities, however, leads to a slight decrease of the score but stagnates to a roughly constant value, as the number of points added increases. This indicates that the calculation of the distortions is fairly tolerant to the presence of noise or wrong elements in the EBSD speckle. However, missing points are more detrimental.

\subsection{The case of $\gamma- \gamma '$ nickel-base superalloys: comparison with other mapping techniques}
\label{subsec:gamma_gammaprime}

The $\gamma$ and $\gamma '$ phases of nickel-based superalloys are well known to be challenging to differentiate, on the single basis of their diffraction patterns. Methods like conventional EBSD with an indexation of phases based on the Hough transform are not efficient. Other indexation techniques have been proposed, such as the dictionary approach by De Graef \citep{Chen2015}. In this approach, the collected patterns are compared to a dictionary of synthetically generated patterns. This technique enables the indexation of patterns with a better angular resolution \citep{Ram2017} as well as the differentiation of phases in complex alloys \citep{Ram2018} and geological materials \citep{Marquardt2017}. However, it can be computationally expensive and -this far- has not been able to differentiate $\gamma$ and $\gamma '$ phases. Combined EDS-EBSD methods have been used in the last years but suffer from long indexing times (and so, increased beam drifting), post-processing times and limited resolution \cite{Charpagne2016}. Using a completely different technique but a similar approach to dictionary indexing, with ion channeling imaging, the iCHORD method of Langlois et al. \cite{Langlois2015} has been shown to enable the differentiation of phases \cite{Langlois2016}. The channeling contrast is used to obtain information about the orientation. On the other hand, the secondary ion signals from $\gamma$ and $\gamma '$ phases are sufficiently contrasted to enable the identification of phases. In a recent comparative study, Vernier et al. \cite{Vernier2018} have shown that this method enables a more accurate mapping than combined EDX-EBSD methods, since it enables to identify precipitates of size down to 150 nm. The scanning speed is also 3 times faster than EDS-EBSD methods. The time necessary for the acquisition and indexation of the patterns however, is typically similar than that of the combined EDS-EBSD method. With a post-processing time of about 15 to 30 minutes, the multimodal data recombination process using the CMA-ES optimizer appears to be a competitive alternative to the other methods proposed in the literature. The only constraint is the acquisition of a BSE image of sufficient quality (in terms of contrast), and its precise segmentation. Indeed, in this method, the size of the smallest objects resolved depends not only on the step size of the EBSD map but also directly on the quality of the segmentation of the BSE image. Assuming a good segmentation, the spatial resolution of this technique is about a few pixels, which is better than EDS-EBSD methods, and comparable to the iCHORD method. It can also be easily implemented in any SEM equipped with a conventional EBSD detector.

\section{Conclusion and perspectives}
\label{sec:conclusion}

A new method for the compensation of the distortions and improved phase differentiation in EBSD data has been developed. The principle consists in using an electron image of the same area than the EBSD data, taken at a 0$\degree$ tilt angle, and using it as a reference. Similar features are segmented out of the EBSD data and electron image. Then, the CMA Evolutionary Strategy is applied in order to match the speckles. The goodness of the superposition is measured by a score, which ranges from 0 to 1. Crystallographic and phase data are then recombined in a new EBSD file. This new file has the same grid than the initial EBSD file.\newline
This method has been applied successfully to nickel-based superalloys and a titanium alloy. Assuming a precise segmentation of the phases on the electron image, this method can reach a precision of a couple of pixels over broad areas, despite important drift phenomena. It can be used on any EBSD dataset, as long as two speckles of similar features can be generated. It is relatively tolerant to inaccuracies and noise in the EBSD speckle. Full automation and short computation times make it a competitive post-processing method, when compared to other options proposed in the literature.

\section*{Acknowledgements}
The authors gratefully acknowledge a Vannevar Bush Fellowship, ONR Grant N00014-18-1-3031. \newline
We acknowledge the following agencies for research funding and computing support: CHISTERA IGLU and CPER Nord-Pas de Calais/FEDER DATA Advanced data science and technologies 2015-2020
Henry Proudhon from Mines ParisTech is acknowledged for prolific conversations and exchange of ideas. 
Olivier Pietquin, J\'er\'emie Mary and Philippe Preux from Inria are acknowledged for insightful conversations. 
Xuedong Shang from INRIA is acknowledged for his various comments on CMA-ES. 
McLean Echlin from the University of Santa Barbara is acknowledged for providing the Ti-6Al-4V sample. 
Mickael Kirka from the Oak Ridge National Laboratory is acknowledged for providing the additive manufactured Inconel 718 sample. 
The Carlton Forge Works company (PCC Corporation) is acknowledged for providing the Rene 65 material.

\newpage
\section*{References}
\bibliography{main.bib}

\begin{thebibliography}{37}
\providecommand{\natexlab}[1]{#1}
\providecommand{\url}[1]{\texttt{#1}}
\providecommand{\urlprefix}{URL }
\expandafter\ifx\csname urlstyle\endcsname\relax
  \providecommand{\doi}[1]{doi:\discretionary{}{}{}#1}\else
  \providecommand{\doi}[1]{doi:\discretionary{}{}{}\begingroup
  \urlstyle{rm}\url{#1}\endgroup}\fi
\providecommand{\bibinfo}[2]{#2}

\bibitem[{Suzuki(2013)}]{Suzuki2013}
\bibinfo{author}{S.~Suzuki}, \bibinfo{title}{{Features of Transmission EBSD and
  its Application}}, \bibinfo{journal}{JOM}
  \bibinfo{volume}{65}~(\bibinfo{number}{9}) (\bibinfo{year}{2013})
  \bibinfo{pages}{1254--1263}, ISSN \bibinfo{issn}{1047-4838},
  \doi{\bibinfo{doi}{10.1007/s11837-013-0700-6}}.

\bibitem[{Echlin et~al.(2015)Echlin, Straw, Randolph, Filevich, and
  Pollock}]{Echlin2015a}
\bibinfo{author}{M.~P. Echlin}, \bibinfo{author}{M.~Straw},
  \bibinfo{author}{S.~Randolph}, \bibinfo{author}{J.~Filevich},
  \bibinfo{author}{T.~M. Pollock}, \bibinfo{title}{{The TriBeam system:
  Femtosecond laser ablation in situ SEM}}, \bibinfo{journal}{Materials
  Characterization} \bibinfo{volume}{100} (\bibinfo{year}{2015})
  \bibinfo{pages}{1--12}, ISSN \bibinfo{issn}{1044-5803},
  \doi{\bibinfo{doi}{10.1016/J.MATCHAR.2014.10.023}}.

\bibitem[{Rowenhorst et~al.(2006)Rowenhorst, Gupta, Feng, and
  Spanos}]{Rowenhorst2006}
\bibinfo{author}{D.~Rowenhorst}, \bibinfo{author}{A.~Gupta},
  \bibinfo{author}{C.~Feng}, \bibinfo{author}{G.~Spanos}, \bibinfo{title}{{3D
  Crystallographic and morphological analysis of coarse martensite: Combining
  EBSD and serial sectioning}}, \bibinfo{journal}{Scripta Materialia}
  \bibinfo{volume}{55}~(\bibinfo{number}{1}) (\bibinfo{year}{2006})
  \bibinfo{pages}{11--16}, ISSN \bibinfo{issn}{13596462},
  \doi{\bibinfo{doi}{10.1016/j.scriptamat.2005.12.061}},
  \urlprefix\url{http://linkinghub.elsevier.com/retrieve/pii/S135964620600039X}.

\bibitem[{Spanos et~al.(2008)Spanos, Rowenhorst, Lewis, and
  Geltmacher}]{Spanos2008}
\bibinfo{author}{G.~Spanos}, \bibinfo{author}{D.~Rowenhorst},
  \bibinfo{author}{A.~Lewis}, \bibinfo{author}{A.~Geltmacher},
  \bibinfo{title}{{Combining Serial Sectioning, EBSD Analysis, and Image-Based
  Finite Element Modeling}}, \bibinfo{journal}{MRS Bulletin}
  \bibinfo{volume}{33}~(\bibinfo{number}{06}) (\bibinfo{year}{2008})
  \bibinfo{pages}{597--602}, ISSN \bibinfo{issn}{0883-7694},
  \doi{\bibinfo{doi}{10.1557/mrs2008.124}}.

\bibitem[{Lin et~al.(2010)Lin, Godfrey, Jensen, and Winther}]{Lin2010}
\bibinfo{author}{F.~Lin}, \bibinfo{author}{A.~Godfrey}, \bibinfo{author}{D.~J.
  Jensen}, \bibinfo{author}{G.~Winther}, \bibinfo{title}{{3D EBSD
  characterization of deformation structures in commercial purity aluminum}},
  \bibinfo{journal}{Materials Characterization}
  \bibinfo{volume}{61}~(\bibinfo{number}{11}) (\bibinfo{year}{2010})
  \bibinfo{pages}{1203--1210}, ISSN \bibinfo{issn}{1044-5803},
  \doi{\bibinfo{doi}{10.1016/J.MATCHAR.2010.07.013}}.

\bibitem[{Calcagnotto et~al.(2010)Calcagnotto, Ponge, Demir, and
  Raabe}]{Calcagnotto2010}
\bibinfo{author}{M.~Calcagnotto}, \bibinfo{author}{D.~Ponge},
  \bibinfo{author}{E.~Demir}, \bibinfo{author}{D.~Raabe},
  \bibinfo{title}{{Orientation gradients and geometrically necessary
  dislocations in ultrafine grained dual-phase steels studied by 2D and 3D
  EBSD}}, \bibinfo{journal}{Materials Science and Engineering: A}
  \bibinfo{volume}{527}~(\bibinfo{number}{10-11}) (\bibinfo{year}{2010})
  \bibinfo{pages}{2738--2746}, ISSN \bibinfo{issn}{0921-5093},
  \doi{\bibinfo{doi}{10.1016/J.MSEA.2010.01.004}}.

\bibitem[{Holzer and Cantoni(2011)}]{Holzer2011}
\bibinfo{author}{L.~Holzer}, \bibinfo{author}{M.~Cantoni},
  \bibinfo{title}{{Review of FIB tomography}}, in:
  \bibinfo{booktitle}{Nanofabrication using focused ion and electron beams :
  principles and applications}, \bibinfo{publisher}{Oxford University Press},
  ISBN \bibinfo{isbn}{9780199734214}, \bibinfo{pages}{410--435},
  \bibinfo{year}{2011}.

\bibitem[{Polonsky et~al.(2018)Polonsky, Echlin, Lenthe, Dehoff, Kirka, and
  Pollock}]{Polonsky2018}
\bibinfo{author}{A.~T. Polonsky}, \bibinfo{author}{M.~P. Echlin},
  \bibinfo{author}{W.~C. Lenthe}, \bibinfo{author}{R.~R. Dehoff},
  \bibinfo{author}{M.~M. Kirka}, \bibinfo{author}{T.~M. Pollock},
  \bibinfo{title}{{Defects and 3D structural inhomogeneity in electron beam
  additively manufactured Inconel 718}}, \bibinfo{journal}{Materials
  Characterization} ISSN \bibinfo{issn}{1044-5803},
  \doi{\bibinfo{doi}{10.1016/J.MATCHAR.2018.02.020}}.

\bibitem[{Nowell and Wright(2004)}]{Nowell2004}
\bibinfo{author}{M.~M. Nowell}, \bibinfo{author}{S.~I. Wright},
  \bibinfo{title}{{Phase differentiation via combined EBSD and XEDS}},
  \bibinfo{journal}{Journal of Microscopy}
  \bibinfo{volume}{213}~(\bibinfo{number}{3}) (\bibinfo{year}{2004})
  \bibinfo{pages}{296--305}, ISSN \bibinfo{issn}{00222720},
  \doi{\bibinfo{doi}{10.1111/j.0022-2720.2004.01299.x}}.

\bibitem[{West and Thomson(2009)}]{West2009}
\bibinfo{author}{G.~West}, \bibinfo{author}{R.~Thomson},
  \bibinfo{title}{{Combined EBSD/EDS tomography in a dual-beam FIB/FEG-SEM}},
  \bibinfo{journal}{Journal of Microscopy}
  \bibinfo{volume}{233}~(\bibinfo{number}{3}) (\bibinfo{year}{2009})
  \bibinfo{pages}{442--450}, ISSN \bibinfo{issn}{00222720},
  \doi{\bibinfo{doi}{10.1111/j.1365-2818.2009.03138.x}}.

\bibitem[{Child et~al.(2012)Child, West, and Thomson}]{Child2012}
\bibinfo{author}{D.~Child}, \bibinfo{author}{G.~West},
  \bibinfo{author}{R.~Thomson}, \bibinfo{title}{{The use of combined
  three-dimensional electron backscatter diffraction and energy dispersive
  X-ray analysis to assess the characteristics of the gamma/gamma-prime
  microstructure in alloy 720Li™}}, \bibinfo{journal}{Ultramicroscopy}
  \bibinfo{volume}{114} (\bibinfo{year}{2012}) \bibinfo{pages}{1--10}, ISSN
  \bibinfo{issn}{03043991},
  \doi{\bibinfo{doi}{10.1016/j.ultramic.2011.11.003}}.

\bibitem[{Charpagne et~al.(2016)Charpagne, Vennegues, Billot, Franchet, and
  Bozzolo}]{Charpagne2016}
\bibinfo{author}{M.-A. Charpagne}, \bibinfo{author}{P.~Vennegues},
  \bibinfo{author}{T.~Billot}, \bibinfo{author}{J.-M. Franchet},
  \bibinfo{author}{N.~Bozzolo}, \bibinfo{title}{{Evidence of multimicrometric
  coherent $\gamma$′ precipitates in a hot-forged $\gamma$–$\gamma$′
  nickel-based superalloy}}, \bibinfo{journal}{Journal of Microscopy}
  \bibinfo{volume}{263}~(\bibinfo{number}{1}), ISSN \bibinfo{issn}{13652818},
  \doi{\bibinfo{doi}{10.1111/jmi.12380}}.

\bibitem[{Payton and Nolze(2013)}]{Payton2013}
\bibinfo{author}{E.~Payton}, \bibinfo{author}{G.~Nolze}, \bibinfo{title}{{The
  Backscatter Electron Signal as an Additional Tool for Phase Segmentation in
  Electron Backscatter Diffraction}}, \bibinfo{journal}{Microscopy and
  Microanalysis} \bibinfo{volume}{19}~(\bibinfo{number}{04})
  (\bibinfo{year}{2013}) \bibinfo{pages}{929--941}, ISSN
  \bibinfo{issn}{1431-9276}, \doi{\bibinfo{doi}{10.1017/S1431927613000305}}.

\bibitem[{Nolze(2007)}]{Nolze2007}
\bibinfo{author}{G.~Nolze}, \bibinfo{title}{{Image distortions in SEM and their
  influences on EBSD measurements}}, \bibinfo{journal}{Ultramicroscopy}
  \bibinfo{volume}{107}~(\bibinfo{number}{2-3}) (\bibinfo{year}{2007})
  \bibinfo{pages}{172--183}, ISSN \bibinfo{issn}{03043991},
  \doi{\bibinfo{doi}{10.1016/j.ultramic.2006.07.003}},
  \urlprefix\url{http://linkinghub.elsevier.com/retrieve/pii/S0304399106001483}.

\bibitem[{Zhang et~al.(2014)Zhang, Elbr{\o}nd, and Lin}]{Zhang2014}
\bibinfo{author}{Y.~Zhang}, \bibinfo{author}{A.~Elbr{\o}nd},
  \bibinfo{author}{F.~Lin}, \bibinfo{title}{{A method to correct coordinate
  distortion in EBSD maps}}, \bibinfo{journal}{Materials Characterization}
  \bibinfo{volume}{96} (\bibinfo{year}{2014}) \bibinfo{pages}{158--165}, ISSN
  \bibinfo{issn}{1044-5803},
  \doi{\bibinfo{doi}{10.1016/J.MATCHAR.2014.08.003}}.

\bibitem[{Kammers and Daly(2013)}]{Kammers2013}
\bibinfo{author}{A.~D. Kammers}, \bibinfo{author}{S.~Daly},
  \bibinfo{title}{{Digital Image Correlation under Scanning Electron
  Microscopy: Methodology and Validation}}, \bibinfo{journal}{Experimental
  Mechanics} \bibinfo{volume}{53}~(\bibinfo{number}{9}) (\bibinfo{year}{2013})
  \bibinfo{pages}{1743--1761}, ISSN \bibinfo{issn}{0014-4851},
  \doi{\bibinfo{doi}{10.1007/s11340-013-9782-x}}.

\bibitem[{Mingard et~al.(2014)Mingard, Jones, and Gee}]{Mingard2014}
\bibinfo{author}{K.~Mingard}, \bibinfo{author}{H.~Jones},
  \bibinfo{author}{M.~Gee}, \bibinfo{title}{{Metrological challenges for
  reconstruction of 3-D microstructures by focused ion beam tomography
  methods}}, \bibinfo{journal}{Journal of Microscopy}
  \bibinfo{volume}{253}~(\bibinfo{number}{2}) (\bibinfo{year}{2014})
  \bibinfo{pages}{93--108}, ISSN \bibinfo{issn}{00222720},
  \doi{\bibinfo{doi}{10.1111/jmi.12100}}.

\bibitem[{Dice(1945)}]{Dice1945}
\bibinfo{author}{L.~R. Dice}, \bibinfo{title}{{Measures of the amount of
  ecologic association between species}}, \bibinfo{journal}{Ecology, Wiley
  Online Library} \bibinfo{volume}{26}~(\bibinfo{number}{3})
  (\bibinfo{year}{1945}) \bibinfo{pages}{297--302}.

\bibitem[{Taha and Hanbury(2015)}]{taha2015metrics}
\bibinfo{author}{A.~A. Taha}, \bibinfo{author}{A.~Hanbury},
  \bibinfo{title}{Metrics for evaluating 3D medical image segmentation:
  analysis, selection, and tool}, \bibinfo{journal}{BMC medical imaging}
  \bibinfo{volume}{15}~(\bibinfo{number}{1}) (\bibinfo{year}{2015})
  \bibinfo{pages}{29}.

\bibitem[{skl(2018)}]{sklearn}
\bibinfo{title}{{Scikit-learn}},
  \bibinfo{howpublished}{\url{http://scikit-learn.org/stable/}},
  \bibinfo{note}{[Online; accessed 23-October-2018]}, \bibinfo{year}{2018}.

\bibitem[{ski(2018)}]{skimage}
\bibinfo{title}{{Scikit-image}},
  \bibinfo{howpublished}{\url{https://scikit-image.org/}},
  \bibinfo{note}{[Online; accessed 23-October-2018]}, \bibinfo{year}{2018}.

\bibitem[{Ib{\'{a}}{\~{n}}ez et~al.(2009)Ib{\'{a}}{\~{n}}ez, Ballerini,
  Cord{\'{o}}n, Damas, and Santamar{\'{i}}a}]{Ibanez2009}
\bibinfo{author}{O.~Ib{\'{a}}{\~{n}}ez}, \bibinfo{author}{L.~Ballerini},
  \bibinfo{author}{O.~Cord{\'{o}}n}, \bibinfo{author}{S.~Damas},
  \bibinfo{author}{J.~Santamar{\'{i}}a}, \bibinfo{title}{{An experimental study
  on the applicability of evolutionary algorithms to craniofacial
  superimposition in forensic identification}}, \bibinfo{journal}{Information
  Sciences} \bibinfo{volume}{179}~(\bibinfo{number}{23}) (\bibinfo{year}{2009})
  \bibinfo{pages}{3998--4028}, ISSN \bibinfo{issn}{00200255},
  \doi{\bibinfo{doi}{10.1016/j.ins.2008.12.029}},
  \urlprefix\url{http://linkinghub.elsevier.com/retrieve/pii/S0020025509000085}.

\bibitem[{Sisniega et~al.(2017)Sisniega, Stayman, Yorkston, Siewerdsen, and
  Zbijewski}]{Sisniega2017}
\bibinfo{author}{A.~Sisniega}, \bibinfo{author}{J.~W. Stayman},
  \bibinfo{author}{J.~Yorkston}, \bibinfo{author}{J.~H. Siewerdsen},
  \bibinfo{author}{W.~Zbijewski}, \bibinfo{title}{{Motion compensation in
  extremity cone-beam CT using a penalized image sharpness criterion}},
  \bibinfo{journal}{Physics in Medicine and Biology}
  \bibinfo{volume}{62}~(\bibinfo{number}{9}) (\bibinfo{year}{2017})
  \bibinfo{pages}{3712--3734}, ISSN \bibinfo{issn}{0031-9155},
  \doi{\bibinfo{doi}{10.1088/1361-6560/aa6869}},
  \urlprefix\url{http://stacks.iop.org/0031-9155/62/i=9/a=3712?key=crossref.84caaff65ba9b2a3f6da8ee49365aca7}.

\bibitem[{Reddy et~al.(2013)Reddy, Panigrahi, Kundu, Mukherjee, and
  Debchoudhury}]{Reddy2013}
\bibinfo{author}{S.~S. Reddy}, \bibinfo{author}{B.~Panigrahi},
  \bibinfo{author}{R.~Kundu}, \bibinfo{author}{R.~Mukherjee},
  \bibinfo{author}{S.~Debchoudhury}, \bibinfo{title}{{Energy and spinning
  reserve scheduling for a wind-thermal power system using CMA-ES with mean
  learning technique}}, \bibinfo{journal}{International Journal of Electrical
  Power {\&} Energy Systems} \bibinfo{volume}{53} (\bibinfo{year}{2013})
  \bibinfo{pages}{113--122}, ISSN \bibinfo{issn}{01420615},
  \doi{\bibinfo{doi}{10.1016/j.ijepes.2013.03.032}},
  \urlprefix\url{http://linkinghub.elsevier.com/retrieve/pii/S014206151300149X}.

\bibitem[{Fateen et~al.(2012)Fateen, Bonilla-Petriciolet, and
  Rangaiah}]{Fateen2012}
\bibinfo{author}{S.-E.~K. Fateen}, \bibinfo{author}{A.~Bonilla-Petriciolet},
  \bibinfo{author}{G.~P. Rangaiah}, \bibinfo{title}{{Evaluation of Covariance
  Matrix Adaptation Evolution Strategy, Shuffled Complex Evolution and Firefly
  Algorithms for phase stability, phase equilibrium and chemical equilibrium
  problems}}, \bibinfo{journal}{Chemical Engineering Research and Design}
  \bibinfo{volume}{90}~(\bibinfo{number}{12}) (\bibinfo{year}{2012})
  \bibinfo{pages}{2051--2071}, ISSN \bibinfo{issn}{02638762},
  \doi{\bibinfo{doi}{10.1016/j.cherd.2012.04.011}},
  \urlprefix\url{http://linkinghub.elsevier.com/retrieve/pii/S0263876212001700}.

\bibitem[{Weber et~al.(2015)Weber, Burnell, Meerts, de~Lange, Dong, Muccioli,
  Pizzirusso, and Zannoni}]{Weber2015}
\bibinfo{author}{A.~C.~J. Weber}, \bibinfo{author}{E.~E. Burnell},
  \bibinfo{author}{W.~L. Meerts}, \bibinfo{author}{C.~A. de~Lange},
  \bibinfo{author}{R.~Y. Dong}, \bibinfo{author}{L.~Muccioli},
  \bibinfo{author}{A.~Pizzirusso}, \bibinfo{author}{C.~Zannoni},
  \bibinfo{title}{{Communication: Molecular dynamics and H-1 NMR of n-hexane in
  liquid crystals}}, \bibinfo{journal}{The Journal of Chemical Physics}
  \bibinfo{volume}{143}~(\bibinfo{number}{1}) (\bibinfo{year}{2015})
  \bibinfo{pages}{011103}, \doi{\bibinfo{doi}{10.1063/1.4923253}}.

\bibitem[{Hansen and Ostermeier(1996)}]{Hansen1996}
\bibinfo{author}{N.~Hansen}, \bibinfo{author}{A.~Ostermeier},
  \bibinfo{title}{{Adapting arbitrary normal mutation distributions in
  evolution strategies: the covariance matrix adaptation}}, in:
  \bibinfo{booktitle}{Proceedings of IEEE International Conference on
  Evolutionary Computation}, \bibinfo{publisher}{IEEE}, ISBN
  \bibinfo{isbn}{0-7803-2902-3}, \bibinfo{pages}{312--317},
  \doi{\bibinfo{doi}{10.1109/ICEC.1996.542381}}, \bibinfo{year}{1996}.

\bibitem[{Hansen and Ostermeier(2001)}]{Hansen2001}
\bibinfo{author}{N.~Hansen}, \bibinfo{author}{A.~Ostermeier},
  \bibinfo{title}{{Completely Derandomized Self-Adaptation in Evolution
  Strategies}}, \bibinfo{journal}{Evolutionary Computation}
  \bibinfo{volume}{9}~(\bibinfo{number}{2}) (\bibinfo{year}{2001})
  \bibinfo{pages}{159--195}, ISSN \bibinfo{issn}{1063-6560},
  \doi{\bibinfo{doi}{10.1162/106365601750190398}}.

\bibitem[{Heaney et~al.(2014)Heaney, Lasonde, Powell, Bond, and
  O'Brien}]{Heaney2014}
\bibinfo{author}{C.~M. Heaney}, \bibinfo{author}{M.~Lasonde},
  \bibinfo{author}{A.~Powell}, \bibinfo{author}{B.~Bond},
  \bibinfo{author}{C.~O'Brien}, \bibinfo{title}{{Development of a New Cast and
  Wrought Alloy (Ren{\'{e}} 65) for High Temperature Disk Applications}}, in:
  \bibinfo{booktitle}{8th International Symposium on Superalloy 718 and
  Derivatives}, \bibinfo{pages}{67--77}, \bibinfo{year}{2014}.

\bibitem[{Kirka et~al.(2017)Kirka, Lee, Greeley, Okello, Goin, Pearce, and
  Dehoff}]{Kirka2017}
\bibinfo{author}{M.~M. Kirka}, \bibinfo{author}{Y.~Lee}, \bibinfo{author}{D.~A.
  Greeley}, \bibinfo{author}{A.~Okello}, \bibinfo{author}{M.~J. Goin},
  \bibinfo{author}{M.~T. Pearce}, \bibinfo{author}{R.~R. Dehoff},
  \bibinfo{title}{{Strategy for Texture Management in Metals Additive
  Manufacturing}}, \bibinfo{journal}{JOM}
  \bibinfo{volume}{69}~(\bibinfo{number}{3}) (\bibinfo{year}{2017})
  \bibinfo{pages}{523--531}, ISSN \bibinfo{issn}{1047-4838},
  \doi{\bibinfo{doi}{10.1007/s11837-017-2264-3}}.

\bibitem[{Chen et~al.(2015)Chen, Park, Wei, Newstadt, Jackson, Simmons, {De
  Graef}, and Hero}]{Chen2015}
\bibinfo{author}{Y.~H. Chen}, \bibinfo{author}{S.~U. Park},
  \bibinfo{author}{D.~Wei}, \bibinfo{author}{G.~Newstadt},
  \bibinfo{author}{M.~A. Jackson}, \bibinfo{author}{J.~P. Simmons},
  \bibinfo{author}{M.~{De Graef}}, \bibinfo{author}{A.~O. Hero},
  \bibinfo{title}{{A Dictionary Approach to Electron Backscatter Diffraction
  Indexing}}, \bibinfo{journal}{Microscopy and Microanalysis}
  \bibinfo{volume}{21}~(\bibinfo{number}{03}) (\bibinfo{year}{2015})
  \bibinfo{pages}{739--752}, ISSN \bibinfo{issn}{1431-9276},
  \doi{\bibinfo{doi}{10.1017/S1431927615000756}}.

\bibitem[{Ram et~al.(2017)Ram, Wright, Singh, and {De Graef}}]{Ram2017}
\bibinfo{author}{F.~Ram}, \bibinfo{author}{S.~Wright},
  \bibinfo{author}{S.~Singh}, \bibinfo{author}{M.~{De Graef}},
  \bibinfo{title}{{Error analysis of the crystal orientations obtained by the
  dictionary approach to EBSD indexing}}, \bibinfo{journal}{Ultramicroscopy}
  \bibinfo{volume}{181} (\bibinfo{year}{2017}) \bibinfo{pages}{17--26}, ISSN
  \bibinfo{issn}{03043991},
  \doi{\bibinfo{doi}{10.1016/j.ultramic.2017.04.016}}.

\bibitem[{Ram and {De Graef}(2018)}]{Ram2018}
\bibinfo{author}{F.~Ram}, \bibinfo{author}{M.~{De Graef}},
  \bibinfo{title}{{Phase differentiation by electron backscatter diffraction
  using the dictionary indexing approach}}, \bibinfo{journal}{Acta Materialia}
  \bibinfo{volume}{144} (\bibinfo{year}{2018}) \bibinfo{pages}{352--364}, ISSN
  \bibinfo{issn}{1359-6454},
  \doi{\bibinfo{doi}{10.1016/J.ACTAMAT.2017.10.069}}.

\bibitem[{Marquardt et~al.(2017)Marquardt, {De Graef}, Singh, Marquardt,
  Rosenthal, and Koizuimi}]{Marquardt2017}
\bibinfo{author}{K.~Marquardt}, \bibinfo{author}{M.~{De Graef}},
  \bibinfo{author}{S.~Singh}, \bibinfo{author}{H.~Marquardt},
  \bibinfo{author}{A.~Rosenthal}, \bibinfo{author}{S.~Koizuimi},
  \bibinfo{title}{{Quantitative electron backscatter diffraction (EBSD) data
  analyses using the dictionary indexing (DI) approach: Overcoming indexing
  difficulties on geological materials}}, \bibinfo{journal}{American
  Mineralogist} \bibinfo{volume}{102}~(\bibinfo{number}{9})
  (\bibinfo{year}{2017}) \bibinfo{pages}{1843--1855}, ISSN
  \bibinfo{issn}{0003-004X}, \doi{\bibinfo{doi}{10.2138/am-2017-6062}}.

\bibitem[{Langlois et~al.(2015)Langlois, Douillard, Yuan, Blanchard,
  Descamps-Mandine, {Van de Moort{\`{e}}le}, Rigotti, and
  Epicier}]{Langlois2015}
\bibinfo{author}{C.~Langlois}, \bibinfo{author}{T.~Douillard},
  \bibinfo{author}{H.~Yuan}, \bibinfo{author}{N.~Blanchard},
  \bibinfo{author}{A.~Descamps-Mandine}, \bibinfo{author}{B.~{Van de
  Moort{\`{e}}le}}, \bibinfo{author}{C.~Rigotti}, \bibinfo{author}{T.~Epicier},
  \bibinfo{title}{{Crystal orientation mapping via ion channeling: An
  alternative to EBSD}}, \bibinfo{journal}{Ultramicroscopy}
  \bibinfo{volume}{157} (\bibinfo{year}{2015}) \bibinfo{pages}{65--72}, ISSN
  \bibinfo{issn}{0304-3991},
  \doi{\bibinfo{doi}{10.1016/J.ULTRAMIC.2015.05.023}}.

\bibitem[{Langlois et~al.(2016)Langlois, Charpagne, Dubail, Douillard, and
  Bozzolo}]{Langlois2016}
\bibinfo{author}{C.~Langlois}, \bibinfo{author}{M.-A. Charpagne},
  \bibinfo{author}{S.~Dubail}, \bibinfo{author}{T.~Douillard},
  \bibinfo{author}{N.~Bozzolo}, \bibinfo{title}{{Ni-based superalloy:
  crystalline orientation mapping and gamma-gamma′ phases discrimination with
  the iCHORD method}}, in: \bibinfo{booktitle}{European Microscopy Congress
  2016: Proceedings}, \bibinfo{publisher}{Wiley-VCH Verlag GmbH {\&} Co. KGaA},
  \bibinfo{address}{Weinheim, Germany}, \bibinfo{pages}{930--931},
  \doi{\bibinfo{doi}{10.1002/9783527808465.EMC2016.6924}},
  \bibinfo{year}{2016}.

\bibitem[{Vernier et~al.(2018)Vernier, Franchet, Lesne, Douillard, Silvent,
  Langlois, and Bozzolo}]{Vernier2018}
\bibinfo{author}{S.~Vernier}, \bibinfo{author}{J.-M. Franchet},
  \bibinfo{author}{M.~Lesne}, \bibinfo{author}{T.~Douillard},
  \bibinfo{author}{J.~Silvent}, \bibinfo{author}{C.~Langlois},
  \bibinfo{author}{N.~Bozzolo}, \bibinfo{title}{{iCHORD-SI combination as an
  alternative to EDS-EBSD coupling for the characterization of
  $\gamma$-$\gamma$′ nickel-based superalloy microstructures}},
  \bibinfo{journal}{Materials Characterization} \bibinfo{volume}{142}
  (\bibinfo{year}{2018}) \bibinfo{pages}{492--503}, ISSN
  \bibinfo{issn}{1044-5803},
  \doi{\bibinfo{doi}{10.1016/J.MATCHAR.2018.06.015}}.

\end{thebibliography}
\bibliographystyle{elsarticle-num-names}

\newpage
\appendix

\section{Update of the parameters in CMA-ES}
\label{App:CMA_update}

Set $m \in \mathbb{R}^{n}$, $\sigma \in \mathbb{R}+, \lambda$ \hfill Input\\
$C = I, p_{c} = 0, p_{\sigma}=0,$ \hfill Initialize parameters\\
$c_{c} \approx 4/n, c_{\sigma} \approx 4/n, c_{1} \approx 2/n^{2}, c_{\mu} \approx \mu_{w}/n^{2}, c_{1} + c_{\mu} \leq  1, d_{\sigma} \approx 1+ \sqrt{\frac{\mu_{w}}{n}}, w_{i=1 ... \lambda}$ such that $\mu_{w} = \frac{1}{\sum_{i=1}^{\mu} w_{i}^{2}} \approx 0.3 \lambda$ \newline
While not terminate, \newline
$x_{i} = m + \sigma y_{i}$, where $y_{i} = \mathcal{N}_{i}(0, C),$ for $i = 1, ..., \lambda$ \hfill Sampling and variation\\
$m \leftarrow \sum_{i=1}^{\mu} w_{i}x_{i:\lambda} = m + \sigma y_{w}$, where $y_{w} = \sum_{i=1}^{\mu} w_{i}x_{i:\lambda}$ \hfill Update mean \\
$p_{c} \leftarrow (1 - c_{c}) p_{c} + \mathbb{1}_{\|p_{\sigma}\| < 1.5\sqrt{n}}\sqrt{1 - (1 - c_{c}^{2})}\sqrt{\mu_{w}}y_{w}$ \hfill Cumulation for the calculation of $C$ \\
$p_{\sigma} \leftarrow (1 - c_{\sigma})p_{\sigma}) + \sqrt{1-(1-c_{\sigma})^{2}} \sqrt{\mu_{w}} C^{-\frac{1}{2}} y_{w}$ \hfill Cumulation for the calculation of $\sigma$ \\
$C \leftarrow (1 - c_{1} - c_{\mu})C + c_{1}p_{c}p_{c}^{T} + c_{\mu} \sum_{i=1}^{\mu} w_{i}y_{i:\lambda}y^{T}_{i:\lambda}$, \hfill Update of the covariance matrix $C$\\
$\sigma \leftarrow \sigma exp(\frac{c_{\sigma}}{d\sigma} (\frac{\|p_{\sigma}\|}{E\| \mathcal{N}(0, I)\|} - 1))$ \hfill Update of the step size $\sigma$.\\

\section{Derivative of the vertical and horizontal components of the distortion function}
\label{App:dFxdFy}
\begin{figure}[h]
\centering
\includegraphics[width=0.45\textwidth]{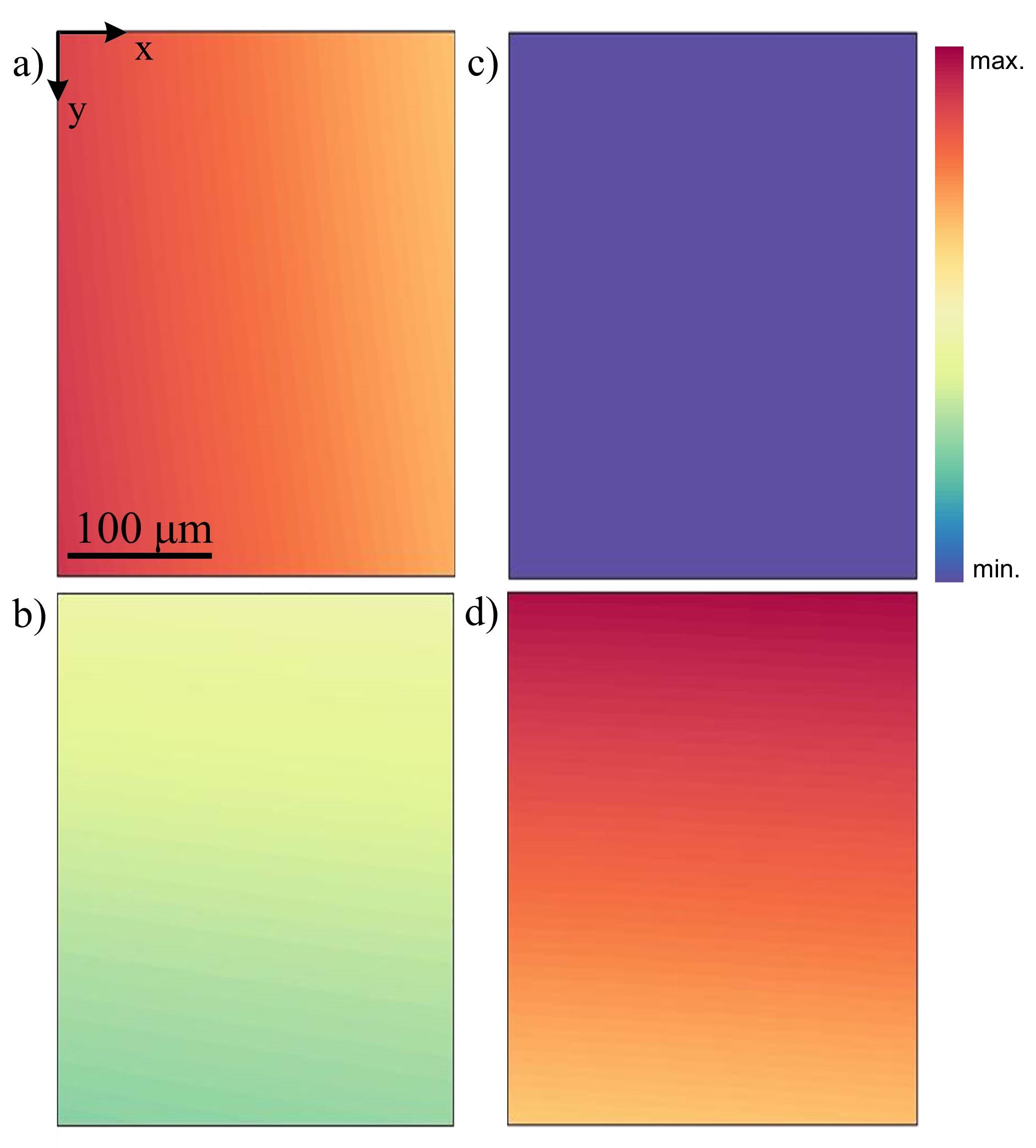}
\caption{Partial derivatives of the vertical and horizontal components of the distortion function along the X and Y coordinates, for the Additive Manufactured Inconel 718 material. a) $\frac{\partial f_{x}}{\partial x}$, b) $\frac{\partial f_{x}}{\partial y}$, c) $\frac{\partial f_{y}}{\partial x}$, d) $\frac{\partial f_{y}}{\partial y}$.}
\label{fig:dFxdFy}
\end{figure}

Figure \ref{fig:dFxdFy} shows 2 sets of maps of the partial derivatives of the vertical and horizontal components of the distortion function reported in fig. \ref{fig:XandYcomponents}. As discussed in section \ref{subsec:versatility}, the variation in color on all four components shows that the error on the location (x',y') varies as the scan progresses. Fig. \ref{fig:dFxdFy} also shows that this variation is not linear. The magnitude of $\frac{\partial f_{x}}{\partial x}$ (\ref{fig:dFxdFy}-a) increases from the left to the right of the map, which means that the error on the x' coordinate increases along each line. This gradient of color is the same along the y component of the map which means that this error remains within the same magnitude from a line to another. The magnitude of this overall gradient remains always much higher than that of $\frac{\partial f_{x}}{\partial y}$  (fig. \ref{fig:dFxdFy}-b), which means that most of the error on the x' coordinates are accumulated along a line. However, the decreasing magnitude of $\frac{\partial f_{x}}{\partial y}$ (fig. \ref{fig:dFxdFy}-c) along the y component of the map, indicates that the error on the x' location decreases from a line to another. this is consistent with a decreasing drift effect as the scan proceeds. As concerns the derivatives of the vertical component $f_{y}$, one can see that most of the error on the y' coordinate occurs along the vertical direction, from a line to another (fig. \ref{fig:dFxdFy}-d). On the other hand, $\frac{\partial f_{y}}{\partial x}$ (fig. \ref{fig:dFxdFy}-c) is fairly constant and low along a line and from a line to another. Those trends are consistent with drift phenomena, where the drift is high in the beginning of the scan, and decreases along a line, and from a line to another.

\end{document}